\pdfoutput=1
\documentclass{vldb}
\usepackage{graphicx}
\usepackage{balance}  

\allowdisplaybreaks

\vldbTitle{Scaling Datalog Processing:  Observations and Techniques}
\vldbAuthors{Zhiwei et al.}
\vldbDOI{}
\vldbVolume{12}
\vldbNumber{xxx}
\vldbYear{2019}

\usepackage{booktabs} 

\usepackage{xspace}
\usepackage{listings}
\usepackage{url}
\usepackage{float}
\usepackage{tabularx,colortbl}
\usepackage{ragged2e}
\usepackage{algorithm}
\usepackage{wrapfig}
\usepackage{lipsum}
\usepackage{float}
\usepackage{graphicx,graphics}
\usepackage[svgnames,table]{xcolor}
\usepackage{algpseudocode}
\usepackage{subcaption}
\usepackage{thmtools, thm-restate}
\usepackage{mathrsfs}
\usepackage{tikz}
\usepackage{comment}
\usepackage{makecell}

\usetikzlibrary{positioning}
\usetikzlibrary{decorations.text}
\usetikzlibrary{decorations.pathmorphing}
\usetikzlibrary{arrows,petri, topaths}
\usepackage[inline]{enumitem}

\newcommand{\aws}[1]{{\color{red}{\noindent\textsf{\textbf{Aws}: #1}}}}

\newcommand{\paris}[1]{{\color{blue}{\noindent\textsf{\textbf{Paris}: #1}}}}

\newcounter{reviewcounter}

\newcommand{\inlineitem}[1][]{%
\ifnum\enit@type=\tw@
    {\descriptionlabel{#1}}
  \hspace{\labelsep}%
\else
  \ifnum\enit@type=\z@
       \refstepcounter{\@listctr}\fi
    \quad\@itemlabel\hspace{\labelsep}%
\fi}
\makeatother

\newcommand{\qstep}{QuickStep\xspace}

\newtheorem{example}{Example}

\newcommand{\cut}[1]{}

\newenvironment{packed_enum}{
\begin{enumerate}
   \setlength{\itemsep}{1pt}
  \setlength{\parskip}{0pt}
   \setlength{\parsep}{0pt}
}
{\end{enumerate}}

\newcommand{\abr}[1]{\textsc{\MakeLowercase{#1}}}
\newcommand{\abrs}[1]{\abr{#1}{\footnotesize{s}}\xspace}

\newcommand{\datalog}{Datalog\xspace}
\newcommand{\datalogsys}{RecStep\xspace}
\newcommand{\sql}{\textsc{sql}\xspace}
\newcommand{\idb}{\textsc{idb}\xspace}
\newcommand{\edb}{\textsc{edb}\xspace}
\newcommand{\rdbms}{\textsc{rdbms}\xspace}
\newcommand{\rdbmss}{\abrs{RDBMS}}
\newcommand{\bddbddb}{\abr{bddbddb}\xspace}
\newcommand{\bdd}{\abr{bdd}\xspace}

\newcommand{\bdds}{\abrs{bdd}\xspace}
\newcommand{\rdds}{\abrs{rdd}\xspace}

\newcommand{\introparagraph}[1]{\vspace{1mm} \noindent \textbf{#1 }}

\newcommand{\PreserveBackslash}[1]{\let\temp=\\#1\let\\=\temp}
\newcolumntype{C}[1]{>{\PreserveBackslash\centering}p{#1}}
\newcolumntype{R}[1]{>{\PreserveBackslash\raggedleft}p{#1}}
\newcolumntype{L}[1]{>{\PreserveBackslash\raggedright}p{#1}}

\newcommand{\cdash}{\mathop{:\!\!-}}

\begin{document}

\title{Scaling-Up In-Memory Datalog Processing:  \\Observations and Techniques}

\numberofauthors{1}
\author{
%
\alignauthor Zhiwei Fan, Jianqiao Zhu, Zuyu Zhang, \\
Aws Albarghouthi,    Paraschos Koutris, Jignesh Patel  \\
\affaddr{University of Wisconsin-Madison} \\
\affaddr{Madison, WI, USA} \\
\email{\{zhiwei, jiangqiao, zuyu, aws, paris, jignesh\}@cs.wisc.edu}
}

\date{\today}

\maketitle

\begin{abstract}
Recursive query processing has experienced a recent resurgence, as a result of its use in many modern application domains, including data integration, graph analytics, security, program analysis, networking and decision making.
Due to the large volumes of data being processed, several research efforts,
across multiple communities, have explored how to scale up recursive queries, typically expressed in Datalog.
Our experience with these tools indicated that their performance does not translate across domains---e.g., a tool design for large-scale graph analytics does not exhibit the same performance on program-analysis tasks,
and vice versa.

As a result, we designed and implemented a general-purpose \datalog\ engine, called \datalogsys, on top of a parallel single-node relational system.
In this paper, we outline the different techniques we use in \datalogsys, and the contribution of each technique to overall performance.
We also present results from a detailed set of experiments comparing \datalogsys with a number of other \datalog systems using both graph analytics and program-analysis tasks,
summarizing pros and cons of existing techniques based on the analysis of our observations.
We show that \datalogsys\ generally outperforms the state-of-the-art parallel Datalog engines on complex and large-scale Datalog program evaluation, by a 4-6X margin.
An additional insight from our work is that we show that it is possible to build a high-performance \datalog system on top of a relational engine, an idea that has been dismissed in past work in this area.
\end{abstract}

\section{Introduction}

Recent years have seen a resurgence of interest from the research and industry community
in the \datalog\ language and its syntactic extensions.
\datalog\ is a recursive query language that extends relational algebra with recursion, and can be
used to express a wide spectrum of modern data management tasks, such as data integration~\cite{DBLP:conf/icdt/FaginKMP03,DBLP:conf/pods/Lenzerini02}, declarative networking~\cite{DBLP:conf/sigmod/LooCGGHMRRS06}, graph analysis~\cite{DBLP:conf/icde/SeoGL13,DBLP:journals/pvldb/SeoPSL13} and program analysis~\cite{whaley04}.
\datalog\ offers a simple and declarative interface to the developer,
while at the same time allowing powerful optimizations that can speed up and scale out evaluation.

Development of Datalog solvers (or engines) has been a subject of study
in both the database community and programming languages community.
The database community independently developed its own tools that evaluate general
 \datalog\ programs,
both in centralized and distributed settings. These include the LogicBlox solver~\cite{green2012logicblox},
as well as distributed, cloud-based engines such as BigDatalog~\cite{BigDatalog}, Myria~\cite{DBLP:journals/pvldb/WangBH15}, and Socialite~\cite{DBLP:journals/pvldb/SeoPSL13}.

In programming languages, it has been observed that a rich class of fundamental static program analyses
can be written equivalently as \datalog\ programs~\cite{reps97,whaley04}.
The programming languages community
has extensively implemented solvers that target all (or a subset) of \datalog.
This line of research has resulted in several \datalog-based tools for program analysis, including
the \bddbddb~\cite{bddbddb}, Souffle~\cite{Souffle}, and more recently Graspan~\cite{Graspan}.

About a year ago, we started exploring the problem of constructing a scalable Datalog engine. We therefore tested \datalog\ engines from across the two communities.
Our experience with these tools indicated that their performance does not translate across domains:
Tools such as LogicBlox and \bddbddb were unable to scale well with large input datasets prevalent in graph analytics.
Even Souffle, the latest and best-performing tool for program analysis tasks, is not well-suited for tasks outside program analysis, such as graph analytics (which also require the language support for
aggregation).
BigDatalog, Myria, and Socialite, on the other hand, can support only the evaluation of simple \datalog\ programs with
limited recursion capabilities ({\em linear recursion} and {\em non-mutual recursion})---however, the majority of program analyses are typically
complex \datalog\ programs with non-linear recursion and mutual recursion.

To address this divide, we asked the following question: {\em Can we design and implement an efficient parallel general-purpose engine that can support a wide spectrum of  \datalog\ programs that target various application domains?}
In this paper, we answer this question positively, by demonstrating how we can implement a fast in-memory parallel
\datalog\ engine using a relational data management system as a backend.
To achieve this goal, we systematically examine the techniques and optimizations necessary to transform
a  na\"ive \datalog\ solver into a highly optimized one that can
practically provide efficient large-scale processing. As a consequence of this work, we also show
that---contrary to anecdotal and empirical evidence~\cite{jordan2016souffle}---it is possible to use effectively an \rdbms\ as a backend for \datalog
through careful consideration of the underlying system issues.

\begin{table*}
\begin{center}
\begin{tabularx}{\textwidth} {C{5cm} C{2cm} C{2cm} C{2cm} C{2cm} C{2cm}}
\toprule
\textbf{} &\  \textbf{Graspan}  &\ \textbf{Bddbddb} &\ \textbf{BigDatalog} &\ \textbf{Souffle}  &\ \textbf{\datalogsys} \\

\textbf{Scale-Up} &\ \textit{yes}   &\ \textit{no}   &\ \textit{yes}  &\ \textit{yes} &\ \textit{yes} \\

\textbf{Scale-Out} &\ \textit{no}   &\ \textit{no}   &\ \textit{yes}  &\ \textit{no} &\ \textit{no} \\

\textbf{Memory Consumption} &\ \textit{low}   &\ \textit{low}   &\ \textit{high}  &\ \textit{medium} &\ \textit{low} \\

\textbf{CPU Utilization} &\ \textit{medium}   &\ \textit{poor}  &\ \textit{high} &\ \textit{medium} &\ \textit{high} \\

\textbf{CPU Efficiency} &\ \textit{low}  &\ - &\ \textit{medium}  &\textit{high} &\  \textit{high} \\

\textbf{Hyperparameters Tuning Requried} &\ \textit{yes (lightweight)}  &\ \textit{yes    (complex)}  &\ \textit{yes (moderate)}  &\ \textit{no} &\  \textit{no} \\

\textbf{Mutual Recursion} &\ \textit{yes} &\ \textit{yes} &\ \textit{no} &\ \textit{yes} &\ \textit{yes} \\

\textbf{Non-Recursive Aggregation} &\ \textit{no} &\ \textit{no} &\ \textit{yes} &\ \textit{yes} &\ \textit{yes} \\

\textbf{Recursive Aggregation} &\ \textit{no}  &\ \textit{no} &\ \textit{yes} &\ \textit{no} &\ \textit{yes} \\

\bottomrule
\end{tabularx}
\caption{Summary of Comparison Between Different Systems: Here we summarize the comparison of different systems in terms of different aspects. Among these aspects, 
\textit{CPU Efficiency} is defined the as the reciprocal of the product of the overall performance (runtime) of the system supporting multi-core computation and the number of CPU cores given for computation - the greater number suggests higher CPU efficiency.
We present the formal definition of CPU efficiency and different systems' CPU efficiency on a few representative workloads in the appendix. The meaning of other aspects should be intuitive as their names suggest.} 
\label{tab:datalog_sys_summary}
\end{center}
\end{table*}

\introparagraph{Our Contribution.}


In this paper, we describe the design and implementation of \datalogsys, a \datalog\ engine that is built based on our observation and analysis of a set of popular Datalog systems. \datalogsys is built on top of \qstep~\cite{quickstep},
which is a single-node in-memory parallel \rdbms and it supports a language extension of
pure \datalog\ with both stratified negation and aggregation, a language fragment that can express a
wide variety of data processing tasks.


 In more detail, we make the following contributions:

\begin{packed_enum}

\item \textbf{Algorithms, data structures, and optimizations:} We systematically study the challenges of building a recursive query processing engine on top of a parallel \rdbms, and propose a set of techniques that solve them. Key techniques include $(i)$ a lightweight way to enable query re-optimization at every recursive step, $(ii)$ careful scheduling of the queries issued to the \rdbms\  in order to maximize resource utilization, and $(iii)$ the design of specialized high-performance algorithms  that target the bottleneck operators of recursive query processing (set difference, deduplication).
We also propose a specialized technique for graph analytics that can compress the intermediate data through
the use of a bit-matrix to reduce memory usage.
We show how the combination of the above techniques can lead to a high-performance \datalog evaluation engine.

\item \textbf{Implementation and evaluation:}
We demonstrate that it is feasible to build a fast general-purpose \datalog\ engine using an \rdbms\ as a
backend. We experimentally show that \datalogsys\ can efficiently solve large-scale problems in different domains
using a single-node multicore machine with large memory, and also can scale well when given more cores.
 We perform an extensive comparison of \datalogsys\ with several other state-of-the-art \datalog\ engines on a multi-core machine. We consider benchmarks both from the graph analysis domain (transitive closure, reachability, connected components), and the program analysis domain (points-to and dataflow analyses) using both synthetic and real-world datasets.
Our findings show that \datalogsys\ outperforms in almost all the cases the other systems, sometimes even by a factor of 8. In addition, the single-node implementation of \datalogsys\ compares favorably
to cluster-based engines, such as BigDatalog, that use far more resources (more processing power and memory).
With the trend towards powerful (multi-core and large main memory) servers, we thus demonstrate that single node systems may be sufficient for a large class of Datalog workloads. The comparison between different systems considering different aspects is summarized in table \ref{tab:datalog_sys_summary}.
\end{packed_enum}

\introparagraph{Organization.}
In Section~\ref{sec:background}, we present the necessary background for \datalog\ and the algorithms used for its
evaluation. Section 4 gives a summary of the architecture design of \datalogsys, while Section 5 discusses in details
the techniques and optimizations that allow us to obtain high-performant evaluation. Finally, in Section 6 we
present our extensive experimental evaluation.

\section{Related Work}
\label{sec:related}

We now compare and contrast our work with existing Datalog engines.

\introparagraph{Distributed Datalog Engines.}
Over the past few years, there have been several  efforts to develop a scalable evaluation engine for Datalog.
Seo et al.~\cite{DBLP:journals/pvldb/SeoPSL13} presented a distributed engine for a
Datalog variant for social network analysis called Socialite.
Socialite employs a number of techniques to enable distribution and scalability,
including delta stepping, approximation of results, and data sharding.
The notable limitation is Socialite's reliance on user-provided annotations to determine how to decompose data on different machines.
Wang et al.~\cite{DBLP:journals/pvldb/WangBH15} implement a variant of
Datalog on the Myria system~\cite{DBLP:conf/sigmod/HalperinACCKMORWWXBHS14},
focusing mostly on asynchronous evaluation and fault-tolerance. The BigDatalog
system~\cite{BigDatalog} is a distributed Datalog engine built on a modified version of Apache Spark.
A key enabler of BigDatalog is a modified version of \rdds in Apache Spark,
enabling fast set-semantic implementation.
The BigDatalog work has shown superior results to the previously proposed systems which we discussed above, Myria and Socialite.
Therefore, in our evaluation, we focus on BigDatalog for a comparison with distributed implementations.
Comparing our work with the above systems, we focus on creating an optimized system atop a parallel in-memory database, instead of a distributed setting.
The task of
parallelizing Datalog has also been studied in the context of the popular
MapReduce
framework~\cite{DBLP:conf/edbt/AfratiBCPU11,DBLP:conf/edbt/AfratiU12,DBLP:conf/datalog/ShawKHS12}.
Motik et al.~\cite{DBLP:conf/aaai/MotikNPHO14} provide an implementation of parallel Datalog in main-memory multicore systems.


\introparagraph{Datalog Solvers in Program Analysis.}
Static program analysis is traditionally the problem of overapproximating runtime program behaviors.
Since the problem is generally undecidable, program analyses resort to overapproximations of runtime facts of a program.
A large and powerful class of program analyses, formulated as context-free language reachability, has been shown to be equivalent to Datalog evaluation.
Thus, multiple Datalog engines have been built and optimized specifically for program-analysis tasks.
The \bddbddb Datalog solver~\cite{bddbddb} pioneered the use of Datalog in program analysis by employing binary decision diagrams (\bdd) to compactly represent the results of program analysis. The idea is that there is lots redundancy in the results of a program analysis, due to overapproximation, that \bdds help in getting exponential savings.
However, recently a number of Datalog solvers have been used for program analysis that employ tabular representations of data. These include the Souffle solver~\cite{Souffle} and the LogicBlox solver~\cite{green2012logicblox}.
Souffle (which has been shown to outperform LogicBlox~\cite{antoniadis2017porting}) compiles Datalog programs into native, parallel C++ code, which is then compiled and optimized for a specific platform.
All of these solvers do not employ a deep form of parallelism that our work exhibits by utilizing a parallel in-memory database.
The Graspan engine~\cite{Graspan} takes a context-free grammar representation and is thus restricted to binary relations---graphs. However, Graspan employs worklist-based algorithm to parallelize fixpoint computation on a multicore machine.
As we show experimentally, however, our approach can outperform Graspan on its own benchmark set.

\introparagraph{Other Graph Engines.}
By now, there are numerous distributed graph-processing systems,
like Pregel~\cite{malewicz2010pregel} and Giraph~\cite{han2015giraph}.
These systems espouse the \emph{think-like-a-vertex} programming model where one writes operations per graph vertex.
These are restricted to binary relations (graphs); Datalog, by definition, is more general, in that it captures computation over hypergraphs.

\section{Background}
\label{sec:background}

In this section, we provide the background on the syntax and evaluation strategies of
the \datalog\ language. Then, we discuss the language extensions we use in this work.

\subsection{Datalog Basics}

A \datalog\ program $P$ is a finite set of rules.
A {\em rule} is an expression of the following form:
\[ h \cdash p_1, p_2, \dots, p_k. \]
The expressions $h, p_1, \dots, p_k$ are {\em atoms}, i.e., formulas of the form
$R(t_1, \dots, t_\ell)$, where $R$ is a table/relation name (predicate) and each  $t_i$ is a {\em term}
that can be a constant  or a variable. An atom is a {\em fact} (or {\em tuple}) when all $t_i$ are constants.
The atom $h$ is called the {\em head} of the rule, and the atoms $p_1, \dots, p_k$
are called the {\em body} of the rule.
A  rule can be interpreted as a logical implication: if the predicates $p_1, \dots, p_k$ are true, then so is the head $h$.
We assume that rules are always safe: this means that all variables in the head occur in at least one atom
in the body.
We will use the lower case $x,y,z, \dots$ to denote variables, and $a,b,c,\dots$ for constants.

The relations in a \datalog\ program are of two types: \idb and \edb relations.
A relation that represents a base (input) table is called \edb (extensional database); \edb relations can occur only in
the body of a rule. A relation that represents a derived relation is called \idb (intentional database), and must
appear in the head of at least one rule.

\begin{example} \label{ex:tc}
Let us consider the task of computing the transitive closure (TC) of a directed graph.
We represent the directed graph using a binary relation {\tt arc(x,y)}: this
means that there is a directed edge from vertex {\tt x} to vertex {\tt y}.
Transitive closure can be expressed through the following program with two rules:
\begin{align*}
r_1: {\tt tc(x, y)} & \cdash {\tt arc(x, y)}. \\
r_2: {\tt tc(x, y)} & \cdash {\tt tc(x, z), arc(z,y)}.
\end{align*}

In the above \datalog\ program, the relation {\tt arc} is an \edb relation (input), and the relation {\tt tc} is an
\idb relation (output).  The program can be interpreted as follows. The first rule $r_1$ (called the {\em base rule})
initializes the transitive closure by adding to the initially empty relation all the edges of the graph.
The second rule $r_2$ is a {\em recursive rule}, and
produces new facts iteratively: a new fact {\tt tc(a,b)} is added whenever there exists some constant {\tt c} such that {\tt tc(a,c)} is already in the relation (of the previous iteration), and {\tt arc(c,b)} is an edge in the graph.
\end{example}

\introparagraph{Stratification.} Given a \datalog\ program $P$, we construct its {\em dependency graph} $G_P$ as follows: every rule is a vertex, and an edge $(r, r')$ exists in $G_P$ whenever the head of the rule $r$
appears in the body of rule $r'$. A rule is {\em recursive} if it belongs in a directed cycle, otherwise it is called non-recursive. A \datalog\ program is recursive if it contains at least one recursive rule.
For instance, the dependency graph of the program in Example~\ref{ex:tc} contains two vertices $r_1, r_2$,
and two directed edges: $(r_1, r_2)$ and $(r_2, r_2)$. Since the graph has a self-loop, the program is recursive.
A {\em stratification} of $P$ is a partition of the rules into strata, where each stratum contains the rules that belong in the same strongly connected component in the dependency graph $G_P$. The topological ordering of the strongly connected components also defines an ordering in the strata.
In our running example, there exist two strata, $\{r_1\}, \{ r_2\}$.


\subsection{Datalog Evaluation}

\datalog\ is a declarative query language, and hence there are different algorithms that can
be applied to evaluate a \datalog\ program.
Most implementations of \datalog\ use {\em bottom-up} evaluation techniques, which start from the input (\edb) tables,
and then iteratively apply the rules until no more new tuples can be added in the \idb relations, reaching a {\em fixpoint}.
In the {\em na\"ive evaluation} strategy, the rules are applied by using all the facts produced so far.
For our running example, we would initialize the \idb relation {\tt tc} with ${\tt tc}^0 \gets {\tt arc}$.
To compute the $(i+1)$ iteration, we compute ${\tt tc}^{i+1} \gets \pi_{x,y} ({\tt tc}^{i} \Join {\tt arc})$. The evaluation
ends when ${\tt tc}^{i+1} = {\tt tc}^i$.
The na\"ive evaluation strategy has the drawback that the same tuples will be produced multiple times throughout the evaluation.

In the {\em semi-na\"ive evaluation} strategy, at every iteration the algorithm uses only the {\em new} tuples from the previous iteration to generate tuples in the current iteration. For instance, in the running example, at every iteration
$i$, we maintain together with ${\tt tc}^i$ the facts that are generated only in the $i$-th iteration (and not previous iterations), denoted by $\Delta {\tt tc}^i = {\tt tc}^i - {\tt tc}^{i-1}$.
Then, we compute the $(i+1)$ iteration of {\tt tc} as ${\tt tc}^{i+1} \gets \pi_{x,y} (\Delta{\tt tc}^{i} \Join {\tt arc})$.
The running example is an instance of {\em linear recursion}, where each recursive rule contains at most one atom with an \idb. However, many \datalog\ programs, especially in the context of program analysis, contain non-linear recursion, where the body of a rule contains multiple \idb atoms. In this case, the $\Delta$ relations are computed by taking the union of multiple subqueries (for more details see~\cite{AliceBook}).

Semi-na\"ive evaluation can be further sped up by exploiting the stratification of a \datalog\ program: the strata are ordered (from lower to higher according to the topological order), and then each stratum is evaluated sequentially, by considering the \idb relations of prior strata as \edb relations (input tables) in the current stratum.
In our implementation of \datalog, we use the  semi-na\"ive evaluation strategy in combination with stratification.

\subsection{Negation and Aggregation}

In order to enhance the expressiveness of \datalog\ for use in modern applications, we consider two
syntactic extensions: {\em negation} and {\em aggregation}.

\introparagraph{Negation.}
\datalog\ is a monotone language, which means that it cannot express tasks where the output can become smaller when the input grows. However, many tasks are inherently non-monotone. To express these tasks, we extend \datalog\ with a simple form of negation, called {\em stratified negation}. In this extension, an atom can be negated by adding the symbol $\neg$ in front of the atom. However, the use of $\neg$ is restricted syntactically, such that an atom $R(t_1, \dots, t_\ell)$ can be negated in a rule if $(i)$ $R$ is an \edb, or $(ii)$ any rule where $R$ occurs in the head is in a strictly lower stratum.

\begin{example}
Suppose we want to compute the complement of transitive closure, in other words, the vertex pairs that do not belong
in the closure. This task can be expressed by the following \datalog\ program with stratified negation:
\begin{align*}
{\tt tc(x, y)} & \cdash {\tt arc(x, y)}. \\
{\tt tc(x, y)} & \cdash {\tt tc(x, z), arc(z,y)}. \\
{\tt node(x)} &\cdash {\tt arc(x,y)}. \\
{\tt node(y)} & \cdash {\tt arc(x,y)}. \\
{\tt ntc(x,y)} & \cdash {\tt node(x), node(y), \neg tc(x,y)}.
\end{align*}
\end{example}

\introparagraph{Aggregation.}
We further extend \datalog\ with aggregation operators.
To support aggregation, we allow the terms in the head of the rule to be of the form ${\tt AGG}(x, y, \dots)$,
where $x,y, \dots$ are variables in the body of the rule, and {\tt AGG} is an aggregation operator that can be
{\tt MIN}, {\tt MAX}, {\tt SUM}, {\tt COUNT}, or {\tt AVG}. 
We allow aggregation not only in non-recursive rules, but also inside recursion as well, which has been early studied in \cite{lefebvre1992towards}. In the latter case, one
must be careful that the semantics of the \datalog\ program lead to convergence to a fixpoint; in this paper, we
assume that the program given as input always converges (\cite{DBLP:conf/amw/ZanioloYIDSC18} studies
how to test this property).
As an example of the use of aggregation, suppose that we want to compute for each vertex the number of
vertices that are reachable from this vertex. To achieve this, we can simply add to the TC \datalog\ program of Example~\ref{ex:tc} the following rule:
\begin{align*}
r_3: {\tt gtc(x, COUNT(y))} & \cdash {\tt tc(x, y)}.
\end{align*}

\smallskip
We should note here that it is straightforward to incorporate negation and aggregation in the standard
semi-na\"ive evaluation strategy, since our syntactic restrictions guarantee that both are applied only when
there is no recursion. Hence, they can easily be encoded in a \sql query: negation as difference, and
aggregation as group-by plus aggregation.

\section{Architecture}

\begin{figure}[t]
   \centering
   \includegraphics[scale=0.4]{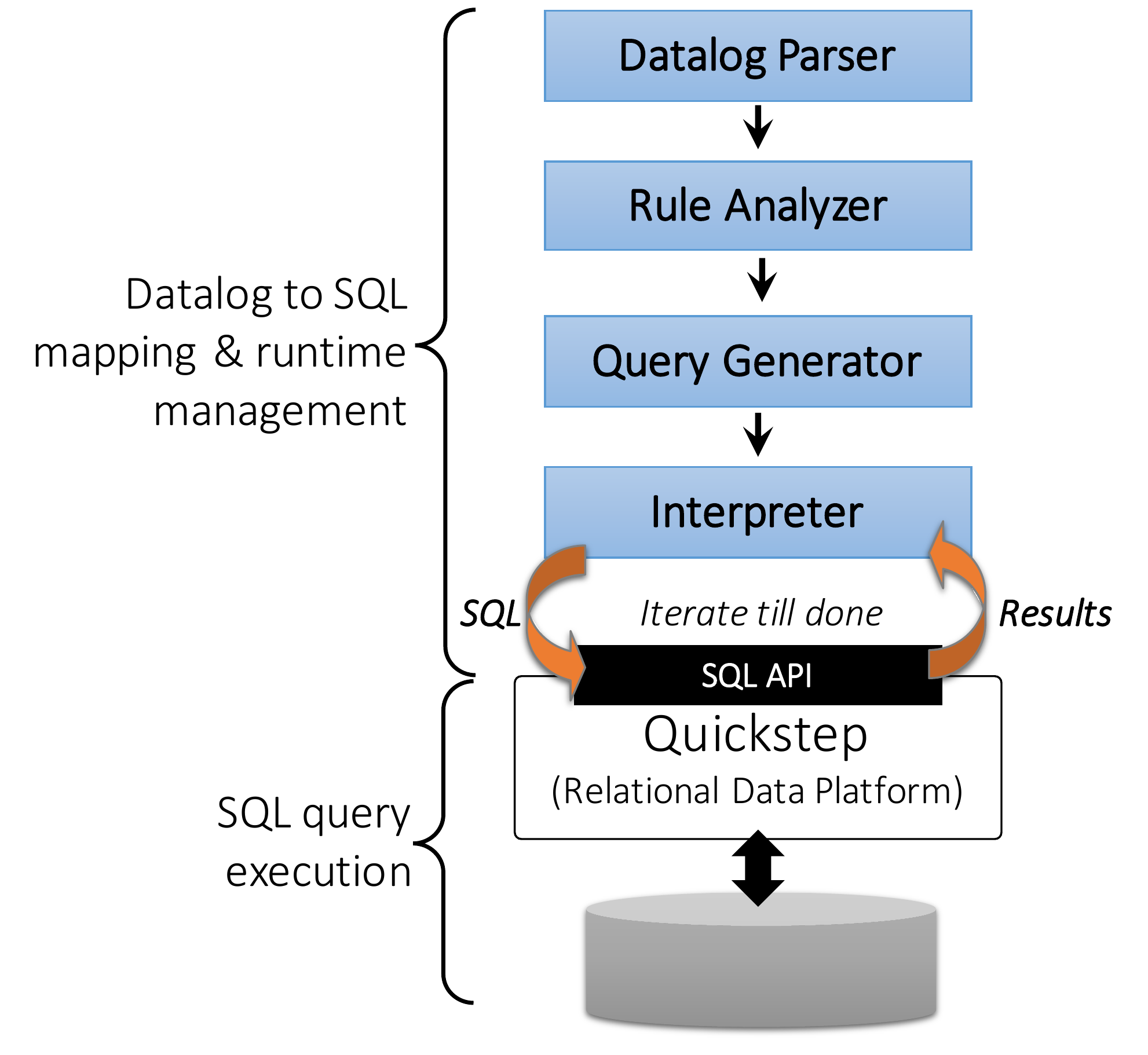}
   \caption{Architectural overview of \datalogsys}
   \label{fig:architecture}
\end{figure}

In this section, we present the architecture of \datalogsys. The core design choice of \datalogsys\ is that, in contrast to  other existing \datalog\ engines, it is built on top of an existing parallel in-memory \rdbms (\qstep). This design enables the use of existing techniques (e.g., indexing, memory management, highly optimized operator implementations) that provide high-performance query execution in a multicore environment. Further, it allows us
to improve the performance by focusing on characteristics specific to \datalog\  evaluation.
On the other hand, as we will discuss in more detail in the next section, it creates the challenging problem of reducing the \rdbms overhead during execution.

\introparagraph{Overview.}
The architecture of our system is summarized in Figure~\ref{fig:architecture}.
The \datalog\ program is read from a {\tt .datalog} file, which, along with the rules of the \datalog\ program,
provides paths for the input and output tables. 
The parsed program is first given as input to the {\em rule analyzer}. The job of the rule analyzer is
to preprocess the program: identify the \idb and \edb relations (and their mapping to input and output tables),
verify the syntactic correctness of the program, and finally construct the dependency graph and stratification.
Next, the {\em query generator} takes the output of the rule analyzer and produces the necessary \sql code to
evaluate each stratum of the \datalog\ program, according to the semi-na\"ive evaluation strategy.
Finally, the {\em interpreter} is responsible for the evaluation of the program. It starts the \rdbms server, creates
the \idb and \edb tables in the database, and is responsible for the loop control for the semi-na\"ive evaluation in each stratum. It also controls the communication and flow of information between the \rdbms server.




\begin{table}
\begin{center}
\begin{tabular} { l l }
\toprule
\textbf{Function} & \textbf{Description} \\
\midrule

$\mathsf{idb}(s)$ &\ returns relations that are heads in stratum $s$ \\
$\mathsf{rules}(R,s)$ &\ returns rules of stratum $s$ with $R$ as head \\
$\mathsf{uieval}(r)$ &\ evaluates all the rules in the set $r$ \\



$\mathsf{analyze}(R)$ &\ call to the \rdbms to collect statistics for $R$\\
$\mathsf{dedup}(R)$ &\ {deduplicates} $R$\\
\bottomrule
\end{tabular}
\caption{Notation used in Algorithm 1}
\label{tab:datalog_eval_alg_notation}
\end{center}
\end{table}


\begin{algorithm}[ht]
\caption{Execution Strategy for \datalog\ program $P$}
\label{alg:datalogeval}
\begin{algorithmic}[1]
\For{\textbf{each} \idb $R$}
\State  $R \gets \emptyset$ \label{alg:init}
\EndFor
\State  // $\mathcal{S}$ is a stratification of $P$
\For{\textbf{each} stratum $s \in \mathcal{S}$}
		\Repeat
			\For{\textbf{each} $R \in \mathsf{idb}(s)$}
				\State $ R_{t} \gets \mathsf{uieval}(\mathsf{rules}(R,s))$ \label{alg:datalogeval:uieval}
				\State $\mathsf{analyze}(R_{t} )$ \label{alg:datalogeval:analyze_1}
				\State $R_{\delta} \gets \mathsf{dedup}(R_{t})$
				\State $\mathsf{analyze}(R_{\delta},R )$ \label{alg:datalogeval:analyze_2}
				\State $\Delta R \gets  R_{\delta}  - R$
				\State $R \gets R \uplus \Delta R $ \label{alg:datalogeval:delta}
			\EndFor
			\If{$s$ is non-recursive} \label{alg:exit}
				\State {\bf break}
			\EndIf
		\Until{$\forall R \in \mathsf{idb}(s)$, $\Delta R = 0$}
	\EndFor
\end{algorithmic}
\end{algorithm}

\introparagraph{Execution.}
We now delve in more detail in how the interpreter executes a \datalog\ program, a procedure
outlined in Algorithm \ref{alg:datalogeval}.

The \datalog\ rules are evaluated in groups and order given by the stratification.
The \idb relations are initialized so that they are empty (line~\ref{alg:init}).
For each stratum, the interpreter enters the control loop for semi-na\"ive evaluation.
Note that in the case where the stratum is non-recursive (i.e., all the rules are non-recursive),
the loop exits after a single iteration (line~\ref{alg:exit}).
In each iteration of the loop, two temporary tables are created for each \idb $R$ in the stratum:
$\Delta R$, which stores only the new facts produced in the current iteration, and a
table that stores the result at the end of the previous iteration.
These tables are deleted right after the evaluation of the next iteration is finished.

The function \textsf{uieval} executes the \sql query that is generated from the query generator
based on the rules in the stratum where the relation appears in the head (more details on the next section).
We should note here that the deduplication
does not occur inside \textsf{uieval}, but in a separate call (\textsf{dedup}). This is achieved in practice by using
{\tt UNION ALL} (simply appending data) instead of  {\tt UNION}.

Finally, we should remark that the interpreter calls the function $\mathsf{analyze}()$ during execution, which tells the backend explicitly to collect statistics on the specified table. As we will see in the next section, $\mathsf{analyze}()$ is a necessary feature to achieve dynamic and
lightweight query optimization.

\section{Optimizations}
\label{sec:opt}

\newcommand{\dsd}{\textsc{dsd}\xspace}
\newcommand{\eost}{\textsc{eost}\xspace}
\newcommand{\oof}{\textsc{oof}\xspace}
\newcommand{\oofNA}{\textsc{oof-na}\xspace}
\newcommand{\oofFA}{\textsc{oof-fa}\xspace}
\newcommand{\fastDedup}{\textsc{fast-dedup}\xspace}

\newcommand{\uie}{\textsc{uie}\xspace}

\begin{figure}[t]
   \centering
   \includegraphics[width=8.5cm]{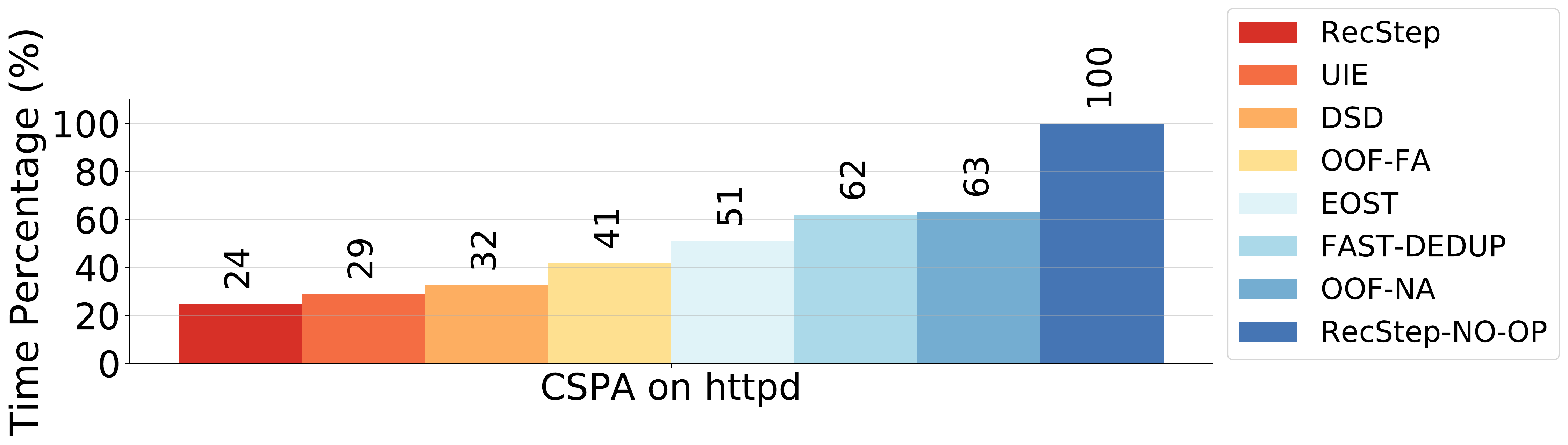}
   \caption{Optimizations for \datalogsys\ :
   	       Here we show the effects of different optimizations by \textit{turning off} each optimization of CSPA analysis on {\tt httpd} dataset . \oofNA means the evaluation uses the same query plan at each iteration; \oofFA means collecting all possible data on updated/new tables (\idb/intermediate results). RecStep-NO-OP means the evaluation with all optimizations turned off. Regarding the total runtime of RecStep-NO-OP (RecStep with all optimizations turned off) as $100\%$ (time percentage), the runtime of evaluation after turning off each optimization is shown as percentage accordingly.
           }
    \label{fig:op}
\end{figure}

\begin{figure}[t]
    \centering
    \begin{subfigure}[b]{0.23\textwidth}
        \centering
        \includegraphics[scale=0.15]{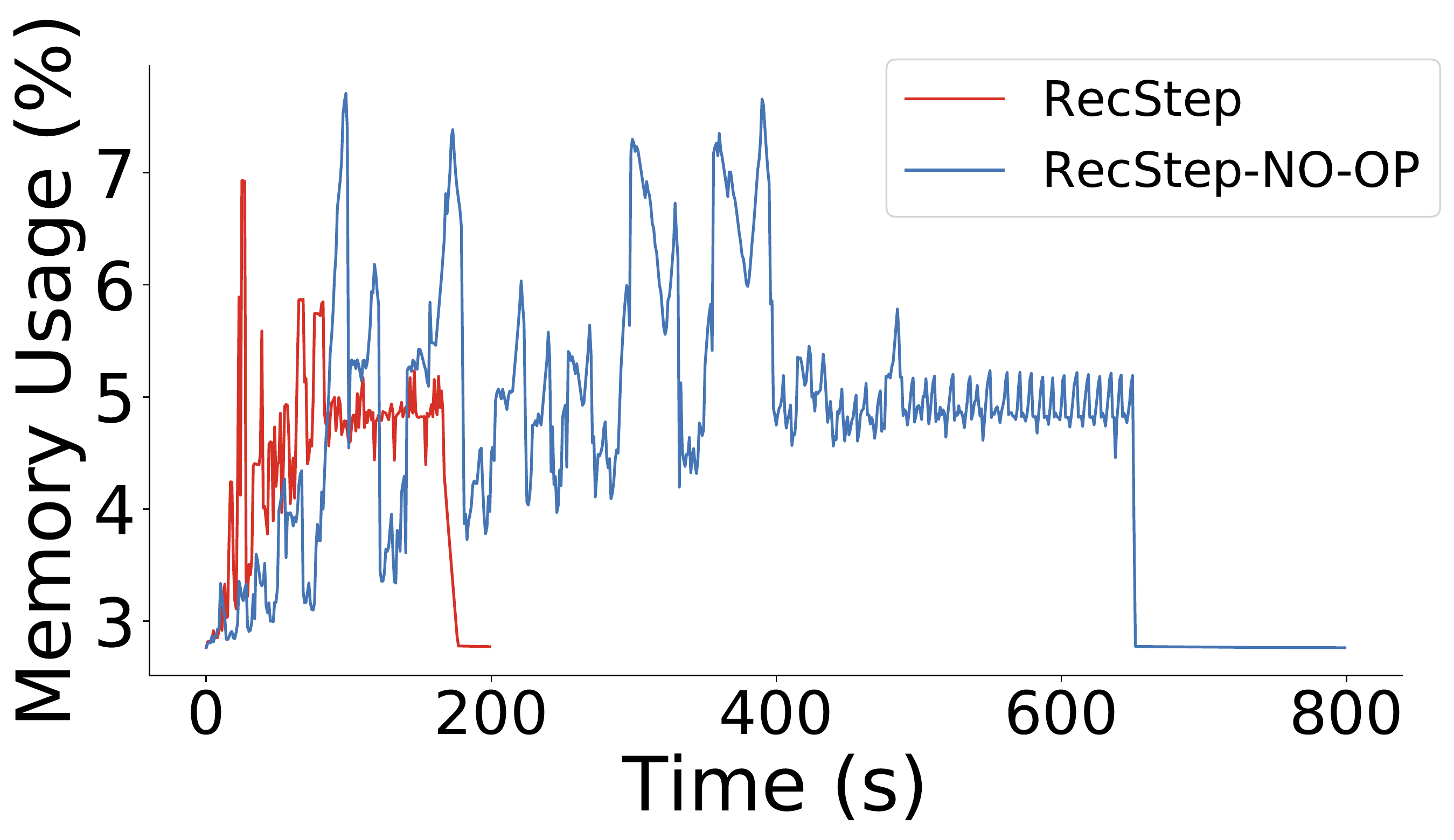}
        \caption{RecStep}
        \label{fig:recstep_full_op_memory}
    \end{subfigure}
    \begin{subfigure}[b]{0.23\textwidth}
        \centering
        \includegraphics[scale=0.15]{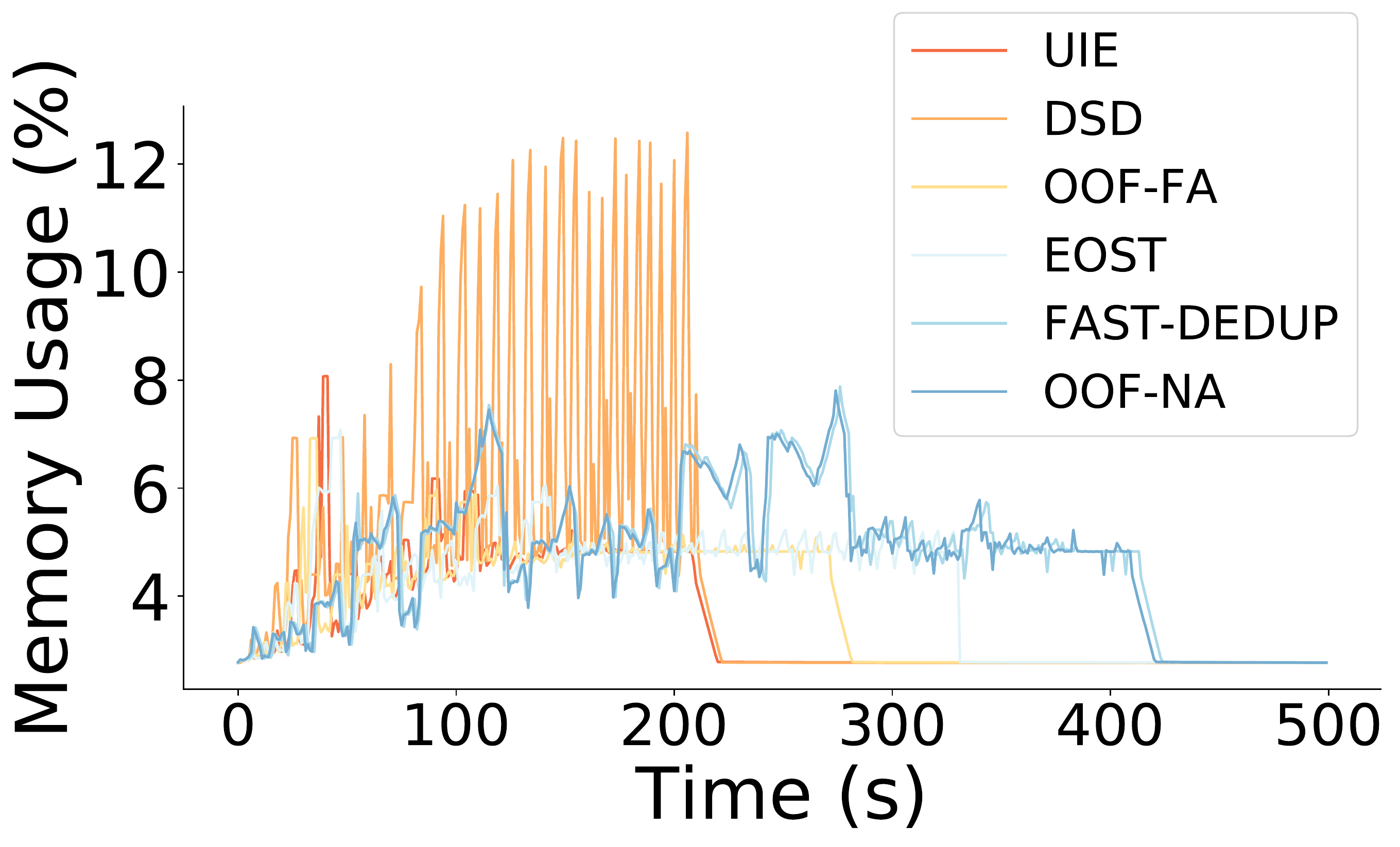}
        \caption{Optimizations}
        \label{fig:recstep_op_only_memory}
    \end{subfigure}
    \caption{Memory Effects of Optimizations: 
         Here Figure \ref{fig:recstep_op_memory} show the memory effect of \textit{turning off} each optimization of CSPA analysis on {\tt httpd}, complimenting Figure \ref{fig:op}.  The memory consumption of RecStep with all optimization turned on/off is shown separately in Figure \ref{fig:recstep_full_op_memory} for comparison purpose.}
    \label{fig:recstep_op_memory}
\end{figure}

This section presents the key optimizations that we have implemented in \datalogsys to speed up performance
and maximize resource utilization (memory and cores).
We consider optimizations in two levels: \datalog-to-\sql-level optimizations and system-level optimizations.

For \datalog-to-\sql-level optimizations,
we study the translation of \datalog\ rules to a set of corresponding \sql queries, so that the evaluation can be done efficiently. This requires careful analysis of the characteristics of the back-end \rdbms (\qstep). Proper translation minimizes the overhead of catalog updates,
selects the optimal algorithms and query plans, avoids redundant computations, and fully utilizes the available parallelism.

In terms of system-level optimizations, we focus on those bottlenecks observed in our experiments that cannot be simply resolved by the translation-level
optimizations. Instead, we modify the back-end system by introducing new specialized data structures, implementing efficient algorithms,
and revising the rules in the query optimizer.

We summarize our optimizations as follows:
\begin{packed_enum}
\item {\em Unified \idb Evaluation} (\uie): different rules and different subqueries inside each recursive rule evaluating the same \idb relation are issued as a single query.
\item {\em Optimization On the Fly} (\oof):  the same set of \sql queries are re-optimized at each iteration considering the change of \idb tables and intermediate results.
\item {\em Dynamic Set Difference} (\dsd): for each \idb table,
the algorithm for performing set-difference to compute $\Delta$ is dynamically chosen at each iteration, by considering the size of \idb tables and intermediate results.
\item{\em Evaluation as One Single Transaction} (\eost): the whole evaluation of a \datalog program is regarded as a single transaction and nothing is committed until the end.
\item{\em Fast Deduplication} (\fastDedup): the specialized implementation of global separate chaining hash table exploiting compact key is used for deduplication to speed up the evaluation and use memory more efficiently 
\end{packed_enum}

Apart from the above list of optimizations, we also provide a specialized technique that can speed up performance on \datalog\ programs that operate on dense graphs called \textsc{pbme}. This technique represents the relation as a bit-matrix, with the goal of minimizing the memory footprint of the algorithm. 
We next detail each optimization;  their effect on runtime and memory is visualized in Figure \ref{fig:op} and Figure \ref{fig:recstep_op_memory}. We present the high level analysis of the time and space effect of each optimization in the appendix.

\subsection{Datalog-to-SQL-level optimizations}

\begin{figure}[t]
   \centering
   \includegraphics[width=6cm]{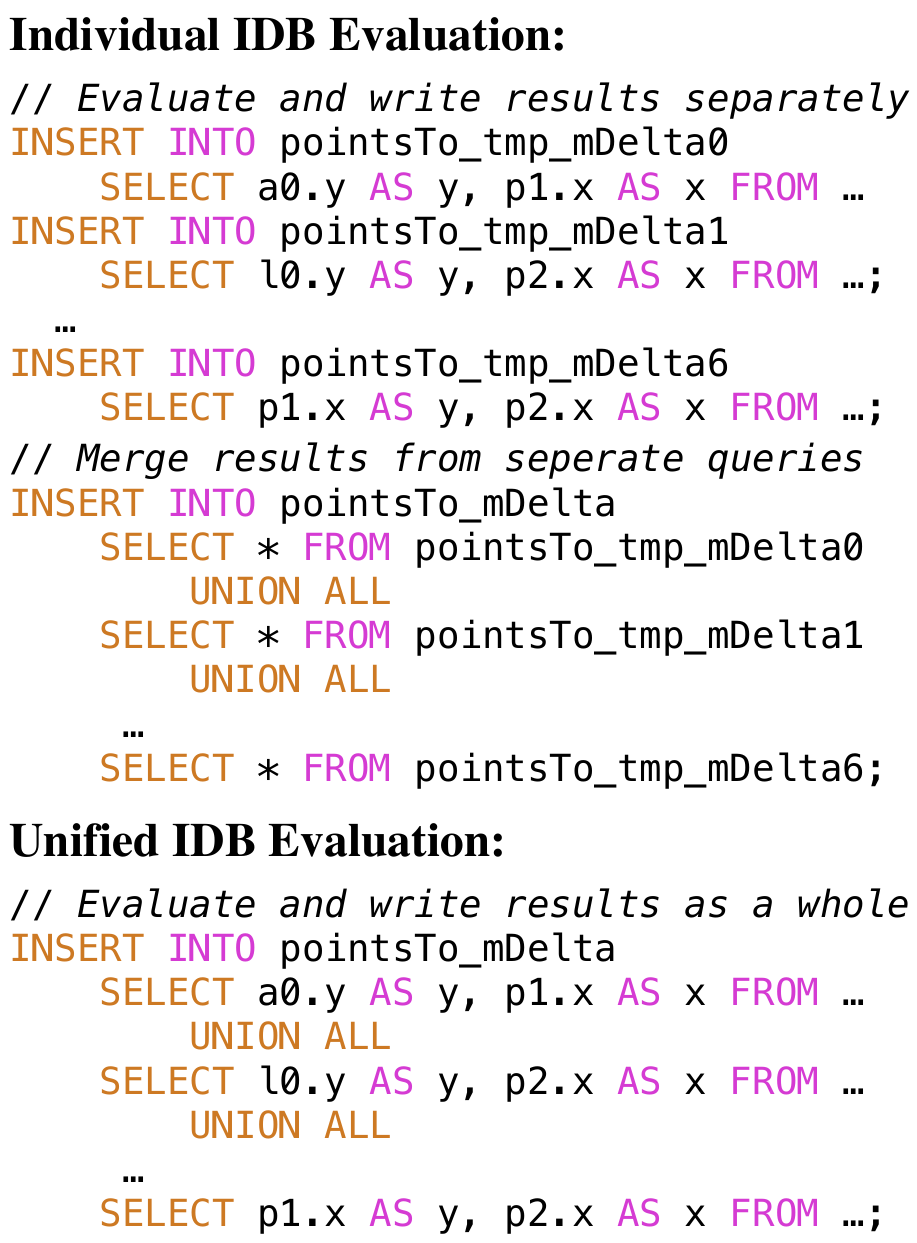}
   \caption{Example \uie in Andersen analysis.}
   \label{fig:uie}
\end{figure}

\introparagraph{Unified IDB Evaluation (UIE).}
For each \idb relation $R$, there can be several rules where $R$ appears in the head of the rule. In addition, for a nonlinear recursive
rule, in which the rule body contains more than one \idb relation, the \idb relation is evaluated by multiple subqueries.
In this case, instead of producing a separate \sql query for each rule and then computing the union of the
intermediate results, the query generator produces a {\em single} \sql query using the {\tt UNION ALL} construct.
We call this method \textit{unified \idb evaluation} (\uie).
Figure~\ref{fig:uie} provides an example of the two different choices for the case of Andersen analysis.

The idea underlying \uie is to  fully utilize all the available resources, i.e., all the cores in a multi-core machine.
\qstep\ does not allow the concurrent execution of \sql\ queries, and hence by grouping the subqueries into a 
single query, we maximize the number of tasks that can be executed in parallel without explicitly considering concurrent multi-task coordination. 
In addition, \uie\ mitigates the overhead incurred by each individual query call, and enables the query optimizer
to jointly optimize the subqueries (e.g., enable cache sharing, or use pipelining instead of materializing intermediate results). The latter point is not specific to \qstep, but generally applicable to any \rdbms\ backend (even ones that support concurrent query processing). 




\introparagraph{Optimization On the Fly (OOF).}
In \datalog\ evaluation, even though the set of queries is fixed across iterations, the input data to
the queries changes, since the \idb relations and the corresponding $\Delta$-relations change at every iteration.
This means that the optimal query plan for each query may be different across different iterations.

For example, in some \datalog programs, the size of $\Delta R$ (Algorithm \ref{alg:datalogeval}) produced in the first few iterations might be much larger than the joining \edb\ table, and thus the hash table should be preferably
built on the \edb\ when performing a join.  However, as the $\Delta R$ produced in later iterations tend to become smaller,  the build side for the hash table should be switched at some point.

In order to achieve optimal performance, it is necessary to re-optimize each query at every iteration (lines~\ref{alg:datalogeval:analyze_1},~\ref{alg:datalogeval:analyze_2} in Algorithm \ref{alg:datalogeval})
by using the latest table statistics from the previous iteration.
However, collecting the statistical data (e.g., size, avg, min, max) of the whole database at every iteration can cause a large overhead, since it may be necessary to perform a full scan of all tables.
To mitigate this issue, our solution is to control precisely at which point which statistical data we collect for the query optimizer, depending on the type of the query.
For instance, before joining two tables, only the size of the two tables is necessary for the optimizer to choose the right side to build the hash table on (the smaller table), as illustrated in the previous example.

In particular, we collect the following statistics:
\vspace{-1ex}
\begin{itemize}
  \setlength{\parskip}{0pt}
  \setlength{\itemsep}{1pt}
\item For {\em deduplication}, the size of the hash table needs to be estimated in order to pre-allocate memory.
Instead of computing the number of distinct values in the table (which could be very expensive), we instead use
a conservative approximation that takes the minimum of the available memory and size of the table.
\item For {\em join processing}, we collect only the number of tuples and the tuple size of the joining tables, if any of the tables is updated or newly created.
\item For {\em aggregation}, we collect statistics regarding the min, max, sum and avg of the tables.
\end{itemize}

The effect of \oof can be seen in Figure~\ref{fig:uie}. Without updating the statistics across the iterations, the running time percentage jumps from $24\%$  to $63\%$ (\oofNA). On the other hand, if we update the full set of statistics, the running time percentage increases to $41\%$ (\oofFA). Thus, this optimization provides a $\sim 2.5\textrm{x}$ speedup in the execution time in for the specific program in the experiment.


\introparagraph{Dynamic Set-Difference (DSD).}
In semi-na\"ive evaluation, the execution engine must compute the \textit{set difference} between the newly evaluated results ($R_{\delta}$)  and the entire recursive relation ($R$) at the end of every iteration, to generate the new  $\Delta{R} \gets R_{\delta}-R$ (line~\ref{alg:datalogeval:delta} in Algorithm \ref{alg:datalogeval}). Since set difference is executed
at every iteration for every \idb in the stratum, it is a computational bottleneck that must be highly optimized.
There exist two different ways we can translate set difference as a \sql\ query.

The first approach (One-Phase Set Difference -- OPSD) simply runs the set difference as a single \sql\ query.  
The default strategy that \qstep\ uses for set difference is to first build a hash table on $R$, and then $R_{\delta}$ probes the hash table to output the tuples of $R_\delta$ that do not match with any tuple in the hash table.
Since the size of $R$ grows at each iteration (recall that \datalog\ is monotone), this suggests that the cost of building the hash table on $R$ will constantly increase for the set difference computation under OPSD.


An alternative approach is to use a translation that we call Two-Phase Set Difference (TPSD).
This approach involves two queries: the first query computes the intersection of the two relations,
$r \gets R \cap R_\delta$. The second query performs set difference, but now between $R$ and $r$ 
(instead of $R_\delta$). Although this approach requires more relational operators, it avoids building a hash
table on $R$. 

We observe that none of the two approaches always dominates the other, since the size of $R$ and $R_{\delta}$ changes at different iterations. Hence, we need to {\em dynamically} choose the best translation at every iteration. To guide this choice, we devise a simple cost model.
In this model, we assign a (per tuple) {\em build cost} $C_b$ during the hash table build phase and a  {\em probe cost} $C_p$ during the hash table probe phase.  Let $\alpha = C_b/C_p $ be the ratio of build to probe cost, $\beta = |R| / |R_{\delta}| $ and $\mu =  |R_{\delta}| /|r|$. We should note here that $\alpha, \beta$ are known before execution, while the intersection size $|r|$ (and hence $\mu$) is unknown. 
Based on the cost model, we can calculate that OPSD is chosen when $\beta \in (0, 1]$ (i.e., when $R$ is the smallest table), and TPSD is chosen if $\beta \geq \frac{2 \alpha}{\alpha - 1}$. When $\beta$ lies in $(1, \frac{2 \alpha}{\alpha - 1})$,  the choice depends on $\mu$, whose value we can approximate using the value of $\mu$ from 
the previous iteration. We present the detailed analysis of the cost model in the appendix.

\subsection{System-level Optimizations} \label{op:sys}

\introparagraph{Evaluation as One Single Transaction (EOST).}
By default, \qstep\  (as well as other \rdbmss) view each query that changes the state of database as a separate transaction. Keeping the default transaction semantics in \qstep\
during evaluation incurs I/O overhead in each iteration due to the frequent insertion in \idb tables, and the creation of tables storing intermediate results.
Such frequent I/O actions are unnecessary,  since we only need to commit the final results at the end of the evaluation. To avoid this overhead, we use the \textit{evaluation as one single transaction} (\eost) semantics. Under these semantics, the data is kept in memory until the fixpoint is reached (when there is enough main memory), and only the final results are written to persistent storage at the end of evaluation.

To achieve \eost,  we slightly modify the kernel code  in \qstep\ to pend the I/O actions until the fixpoint is reached (by default, if some pages of the table are found {\em dirty} after a query execution, the pages are written back to the disk). At the end of the evaluation, a signal is sent to \qstep\ and the data is written to disk. 
For other popular \rdbmss (e.g., PostgreSQL, MySQL, SQL Server), the start and the end of a transaction can be explicitly specified, but this approach is only feasible for a set of queries that are {pre-determined}.\footnote{To fully achieve \eost, transactional databases also need to turn off features such as checkpoint, logging for recovery, etc} 
However, in recursive query processing the issued queries are dynamically generated, and the number of iterations is not known until the fixpoint is reached, which means that similar changes need to be made in these systems to apply \eost.

\begin{figure}[t]
   \centering
   \includegraphics[width=8cm]{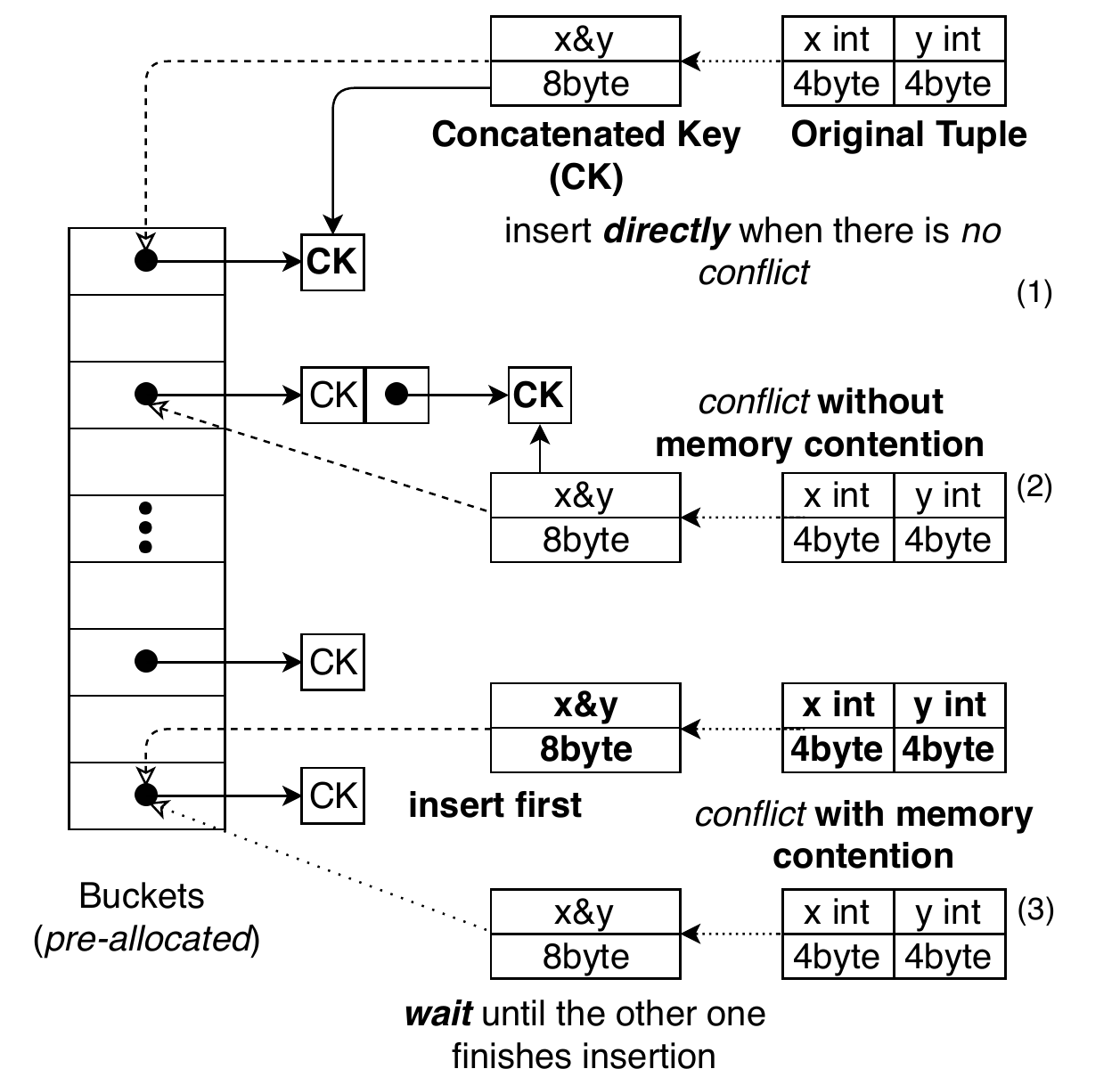}
   \caption{Example of applying fast deduplication algorithm on table with two integer attributes in \datalogsys\ }
   \label{fig:fast_dedup}
\end{figure}

\introparagraph{Fast Deduplication.}
In \datalog\ evaluation, \textit{deduplication} of the evaluated facts is not only necessary for conforming to the {set semantics}, but also helps
to avoid redundant computation.  Deduplication is also a computational bottleneck, since it occurs at every iteration for every \idb\ in the stratum  (line 10 in Algorithm \ref{alg:datalogeval}); hence, it is necessary to optimize its parallel execution.  

To achieve this, we use a specialized \textit{Global Separate Chaining Hash Table} implementation that uses a  \textit{Compact Concatenated Key} (\textsc{ck}), which we call \textsc{cck-gscht}. \textsc{cck-gscht} is a global latch-free hash table built using a compact representation of $\langle key, value \rangle$ pairs, in which tuples from each data partition/block can be inserted \textit{in parallel}. Figure~\ref{fig:fast_dedup} illustrates the deduplication algorithm using an example in which \textsc{cck-gscht} is applied on a table with two integer attributes (src int, dest int).

Based on the approximated number of distinct elements from the query optimizer, we {\em pre-allocate} a list of buckets, where each bucket contains only a pointer.  An important point here is that the number of pre-allocated buckets will be {as large as possible} when there is enough memory, for the purpose of minimizing conflicts in the same bucket, and preventing memory contention. Tuples are assigned to each thread in a round-robin fashion and are inserted in parallel.  Knowing the length of each attribute in the table, a compact \textsc{ck} of fixed size
\footnote{The inputs of \datalog programs are usually integers transformed by mapping the {\em active domain} of the original data (if not integers). Thus the technique can also applied to data where the original type has varied length.}
is constructed for each tuple (8 bytes for two integer attributes as shown in Figure~\ref{fig:fast_dedup}). The compact \textsc{ck} itself contains all information of the original tuple, eliminating the need for explicit $\langle key, value \rangle$ pair representation. Additionally, the key itself is used as the hash value, which saves computational cost and memory space.

\subsection{Parallel Bit-Matrix Evaluation} 

\begin{algorithm}[t]
    \caption{Parallel Bit-Matrix Evaluation of TC
    }
\label{alg:bit_matrix_eval_tc}
\begin{algorithmic} [1]
\State \textbf{Input:}  $arc(x,y)$ - $\edb$ relation, number of threads $k$
\State \textbf{Output:} $tc(x,y)$  - $\idb$ relation
\State // $M_{arc}$: virtual bit-matrix of ${\tt arc(x,y)}$ \label{alg:bit_matrix_eval_tc:construct_bit_matrix_virtual}
\State Construct bit-matrix $M_{tc}$ of ${\tt tc(x,y)}$ 
\State $M_{tc} \gets M_{arc}$
\State Partition the rows of $M_{tc}$ into $k$ partitions  \label{alg:bit_matrix_eval_tc:partition}
\State // the $k$ threads evaluate $k$ partitions in parallel
\For{\textbf{each} row $i$ in partition $p$}
		\State $\delta \gets \{u \mid M_{tc}[i,u] =1 \}$
		\While{$\delta \neq \emptyset$}
			\State $\delta_n \gets \emptyset$
			\For{\textbf{each} $t \in \delta$ } \label{alg:bit_matrix_eval_tc:compute_delta_start} 
				\For{\textbf{each} $j$ s.t. ${M_{arc}}[t,j] = 1$}  \label{alg:bit_matrix_eval_tc:compute_delta} 
				    \If{${M_{tc}}[i,j] = 0$}
				    	\State $\delta_n \gets \delta_n \cup \{j\}$
					\State ${M_{tc}}[i,j] \gets 1$ \label{alg:bit_matrix_eval_tc:delta_update}
				    \EndIf
				\EndFor
			\EndFor \label{alg:bit_matrix_eval_tc:compute_delta_end}
			\State $\delta \gets \delta_n$
		\EndWhile
\EndFor
\end{algorithmic}
\end{algorithm}

\begin{algorithm}[t]
\caption{Parallel Bit-Matrix Evaluation of SG}
\label{alg:bit_matrix_eval_sg}
\begin{algorithmic} [1]
\State \textbf{Input:}  ${\tt arc(x,y)}$ - $\edb$ relation, number of threads $k$
\State \textbf{Output:} ${\tt tc(x,y)}$  - $\idb$ relation
\State // $M_{arc}$: virtual bit-matrix of ${\tt arc(x,y)}$
\State Construct vector index $V_{arc}[x] =  \{y$ $|$ $M_{arc}[x,y] = 1\}$  \label{alg:bit_matrix_eval_sg:index} 
\State Construct $M_{sg}$ bit-matrix of ${\tt sg(x,y)}$ 
\State // ${\tt sg(x,y)} \gets {\tt arc(p,x), arc(p,y), x \gets y.}$
\State // ${\tt arc_1} \leftarrow$ $\textbf{Rename}_{arc_1(x_1,y_1)}({\tt arc})$
\State // ${\tt arc_2} \leftarrow$ $\textbf{Rename}_{arc_2(x_2,y_2)}({\tt arc})$
\State $M_{sg} \gets \Pi_{y_1, y_2}(M_{arc_1}  \bowtie_{x_1=x_2, y_1 \neq y_2} M_{arc_2})$

\State Partition the rows of $M_{sg}$ into $k$ partitions
\State // $k$ threads evaluate $k$ partitions in parallel
\For{\textbf{each} row $i$ in partition $p$}
	\State $\delta \gets \{(i,u) \mid {M_{sg}}[i,u] = 1\}$
	\While{$\delta \neq \emptyset$}
			\State $\delta_n \gets \emptyset$
			\For{\textbf{each} $(a,b) \in \delta$}
				\For{\textbf{each} $q \in {V_{arc}}[a]$}   \label{alg:bit_matrix_eval_sg:use_index1}
					\For{\textbf{each} $p \in {V_{arc}}[b]$} \label{alg:bit_matrix_eval_sg:use_index2}
						\If{${M_{sg}[q,p]} = 0$}
				    			\State $\delta_n \gets \delta_n \cup \{(q,p)\}$ \label{alg:bit_matrix_eval_sg:delta}
							\State $M_{sg}[q,p] \gets 1$  \label{alg:bit_matrix_eval_sg:update}
						\EndIf
					\EndFor
				\EndFor
			\EndFor
			\State $\delta \gets \delta_n $
	\EndWhile
\EndFor
\end{algorithmic}
\end{algorithm}

In our experimental evaluation, we observed that the usage of memory increases drastically during the evaluation of {graph analytics} over dense graphs. By default, \qstep\ uses hash tables for joins between tables, aggregation and deduplication. When the intermediate result becomes very large, the use of hash tables for join processing becomes 
memory-costly. In the extreme, the intermediate results are too big to fit in main memory, and are forced to disk, incurring additional I/O overhead. This can cause in the worst case scenario out-of-memory errors (\textsc{oom}).  Additionally, the materialization cost of large intermediate results can hurt performance.

Besides graph analytics on dense graphs, in many program analyses \idb\ relations, even though they can be large, they are very dense and the active domain of the relational attributes is relatively smaller. This phenomenon is also observed in dense graph analytics, where graphs with a small number of vertices produce results that are a orders of magnitude larger than the inputs.

\begin{figure}[t]
    \centering
    \begin{subfigure}[b]{0.23\textwidth}
        \centering
        \includegraphics[scale=0.14]{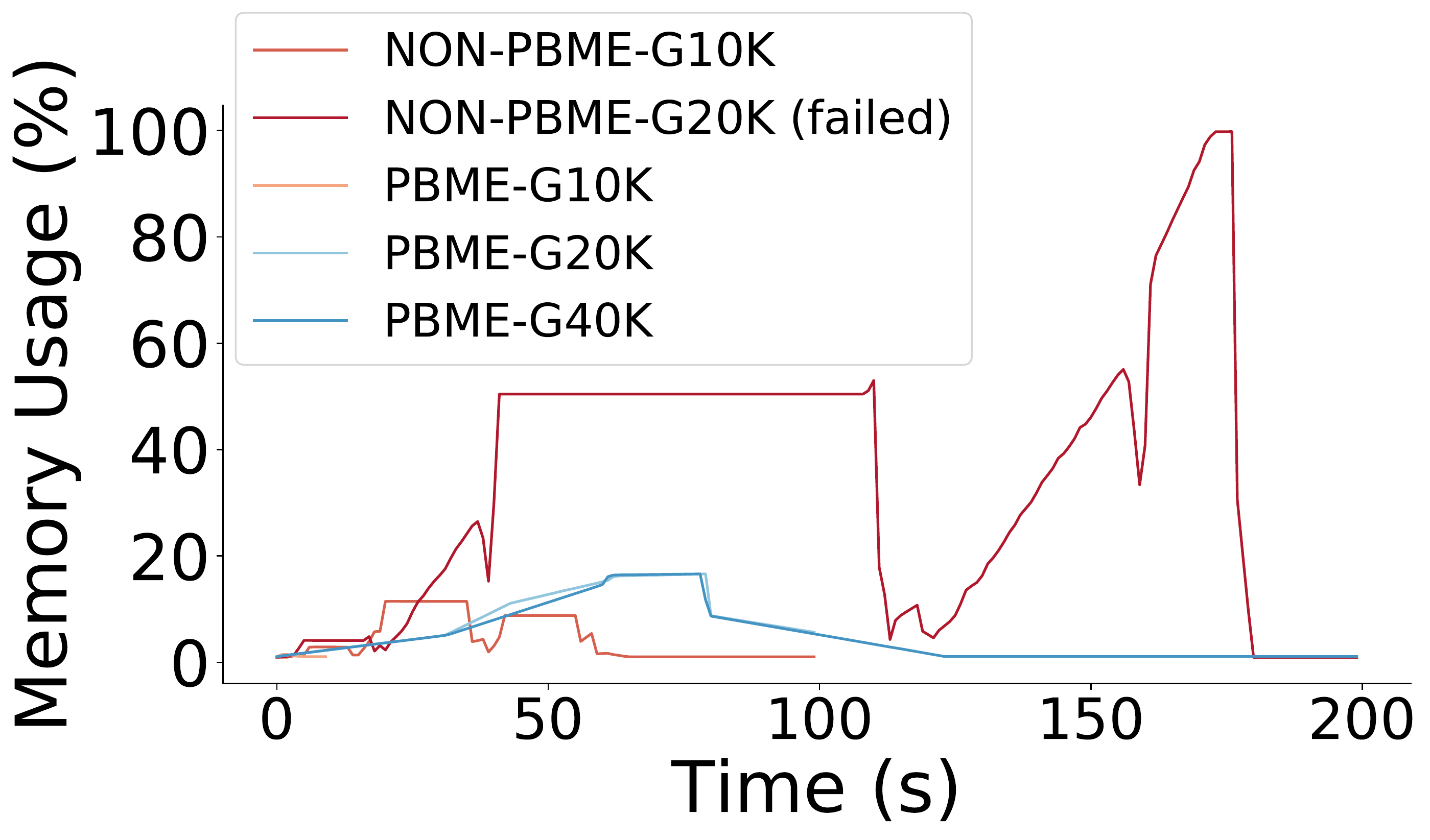}
        \caption{Transitive Closure}
        \label{fig:tc_memory}
    \end{subfigure}
    \begin{subfigure}[b]{0.23\textwidth}
        \centering
        \includegraphics[scale=0.14]{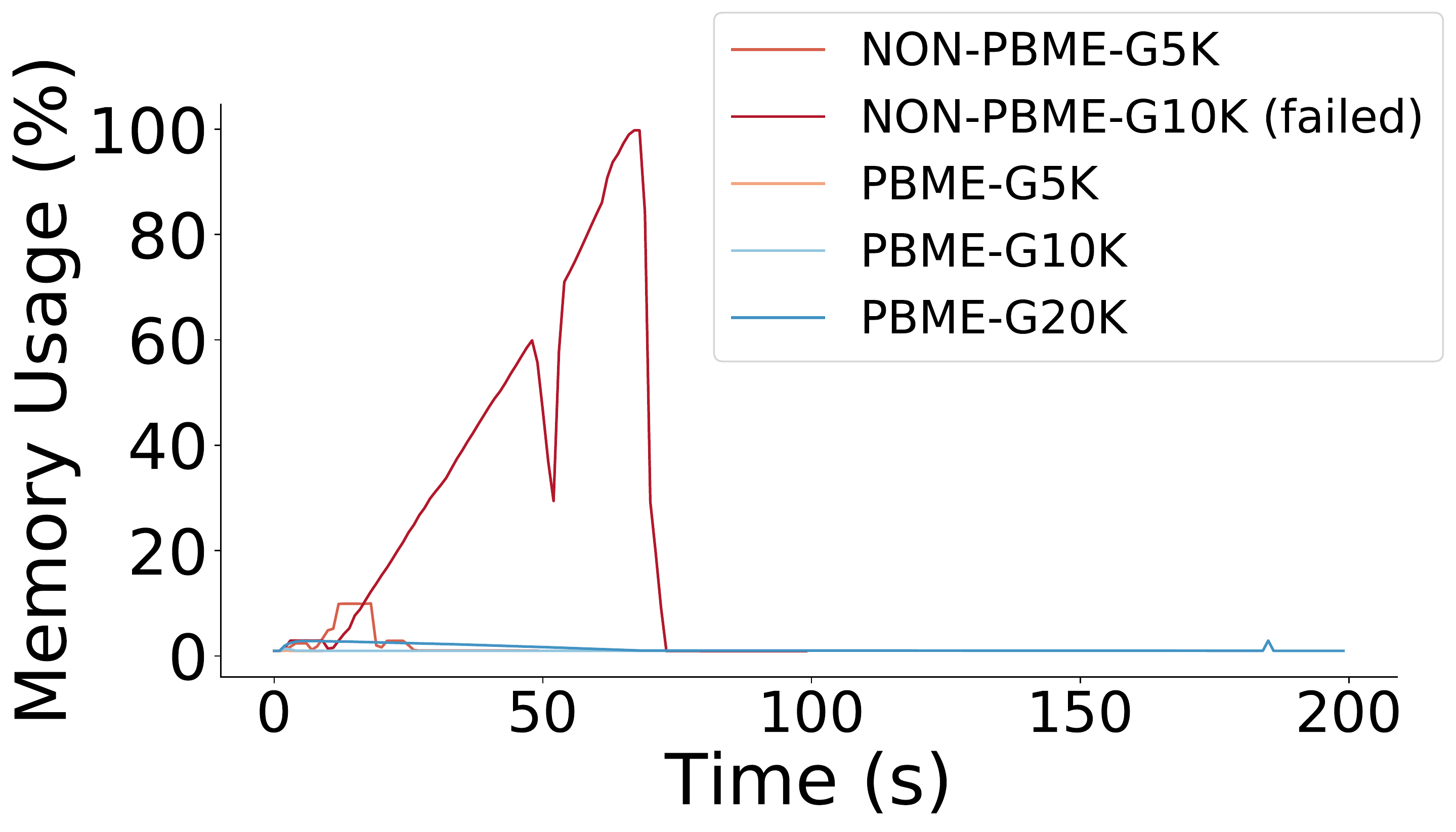}
        \caption{Same Generation}
        \label{fig:sg_memory}
    \end{subfigure}
    \caption{ Memory Saving of PBME on TC and SG)}
    \label{fig:tc_sg_memory_saving}
\end{figure}

Inspired by this observation, we exploit a specialized data structure, called \textit{bit-matrix}, that replaces a hash map during join and deduplication in the case when the graph representing the data is \textit{dense} and has relatively small number of vertices.  This data structure represents the join results in a much more compact way under certain conditions, greatly reducing the memory cost compared to a hash table (The comparison of memory consumption before using PBME and after in TC and SG  is visualized in Figure \ref{fig:tc_sg_memory_saving}). In this paper, we only describe the bit matrix for binary relations, but we note that the technique can be extended to relations of higher arity.
At the same time, we implement new operators directly operating on the {bit-matrix}, naturally merging the join and deduplication into \textit{one single stage} and thus avoid the materialization cost coming from the intermediate results. We
call this technique \textit{Parallel Bit-Matrix Evaluation} (\textsc{pbme}). In our experiments, \textsc{pbme} provided a huge boost in performance for transitive closure (TC) and same generation (SG).

We next describe the {bit-matrix} technique and present two examples in which it is used for evaluation of \textit{TC} and \textit{SG} as outlined in Algorithm \ref{alg:bit_matrix_eval_tc} and Algorithm \ref{alg:bit_matrix_eval_sg}.
Note that we construct a matrix only for each \idb, but for convenience of illustration, we use matrix notation for 
the \edb{s} as well (line~\ref{alg:bit_matrix_eval_tc:construct_bit_matrix_virtual} in Algorithm \ref{alg:bit_matrix_eval_tc} and Algorithm \ref{alg:bit_matrix_eval_sg}).

\smallskip
\introparagraph{The Bit-Matrix Data Structure.} Let $R(x,y)$ be a binary \idb relation, with active domain $\{1, 2, \dots, n\}$
for both attributes.
Instead of representing $R$ as a set of tuples, we represent it as an $n \times n$ bit matrix denoted $M_R$.
If $R(a,b)$ is a tuple in $R$, the bit at the $a$-th row and $b$-th column, denoted ${M_R}[a,b]$ is set to 1, otherwise it is 0. The relation is updated during recursion by setting more bits from 0 to 1.
We decide to build the bit-matrix data structure only if the memory available can fit both the bit matrix, as well as any additional index data structures used during evaluation (e.g., Algorithm \ref{alg:bit_matrix_eval_sg}).


One of the key features of \textsc{pbme} is \textit{zero-coordination}: each thread is only responsible for the partition of data assigned to it and there is no or nearly no coordination needed between different threads. 
Algorithm~\ref{alg:bit_matrix_eval_tc} outlines the \textsc{pbme} for transitive closure.
For the evaluation of TC (Algorithm \ref{alg:bit_matrix_eval_tc}), the rows of the \idb\ bit-matrix $M_{tc}$ are firstly partitioned in a round-robin fashion (line~\ref{alg:bit_matrix_eval_tc:partition}).  For each row $i$ assigned to each thread,  the set $\delta$  stores the new bits (paths starting from $i$) produced at every iteration (line~\ref{alg:bit_matrix_eval_tc:compute_delta_start}-\ref{alg:bit_matrix_eval_tc:compute_delta_end}). For each new bit $t$ produced, the thread searches for all the bits at row $t$ of $M_{arc}$, and computes the new $\delta$ (line~\ref{alg:bit_matrix_eval_tc:compute_delta}-\ref{alg:bit_matrix_eval_tc:compute_delta_end}). 

The \datalog\ program of SG is as follows:
\begin{align*}
{\tt sg(x,y)} & \cdash {\tt arc(p,x), arc(p,y), x != y}.\\
{\tt sg(x,y)} & \cdash  {\tt arc(a,x), sg(a,b), arc(b,y)}.
\end{align*}

\begin{figure}[t]
    \centering
      \begin{subfigure}[b]{0.23\textwidth}
        \centering
        \includegraphics[scale=0.14]{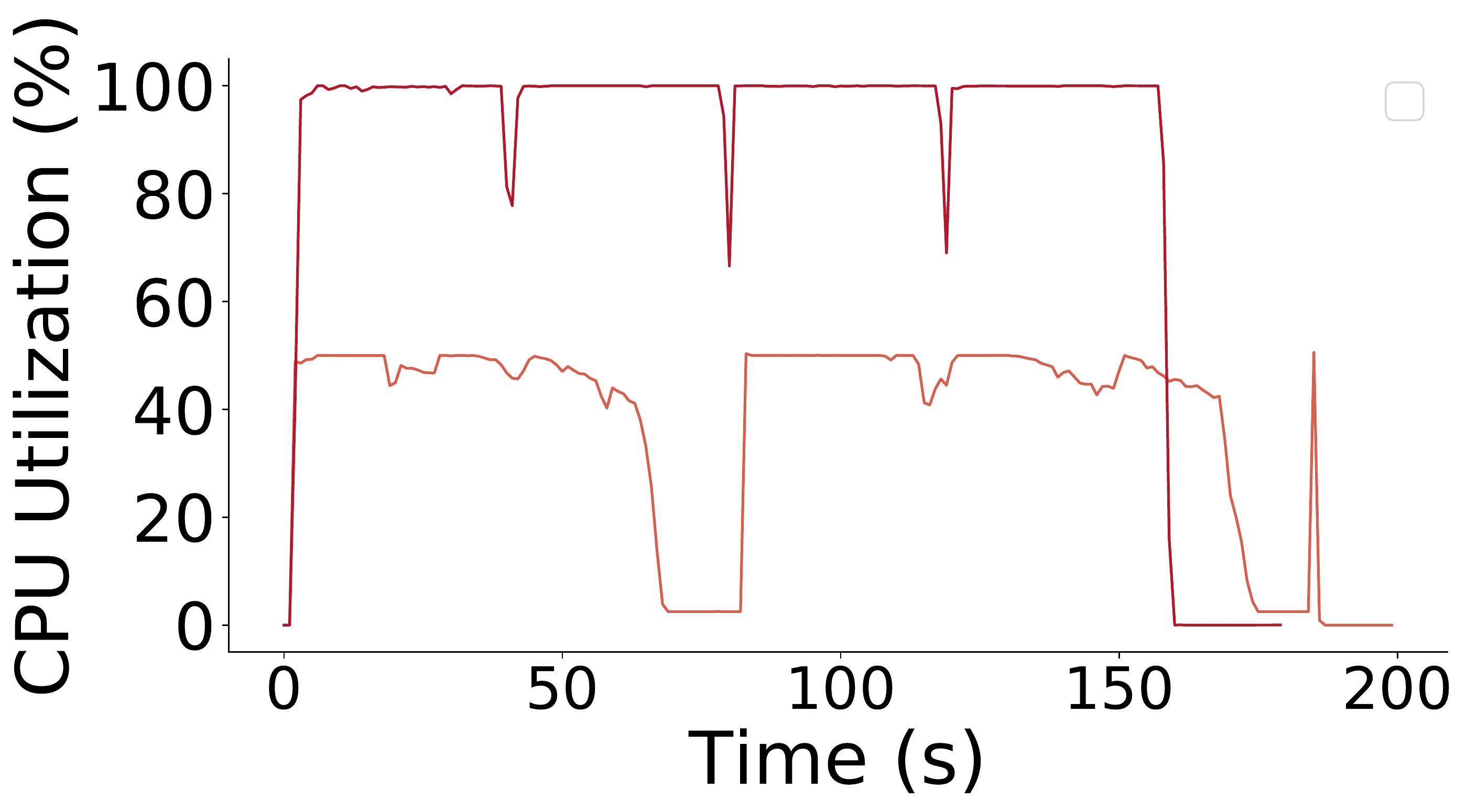}
        \caption{CPU Utilization}
        \label{fig:sg_alt_memory}
    \end{subfigure}
    \begin{subfigure}[b]{0.23\textwidth}
        \centering
        \includegraphics[scale=0.14]{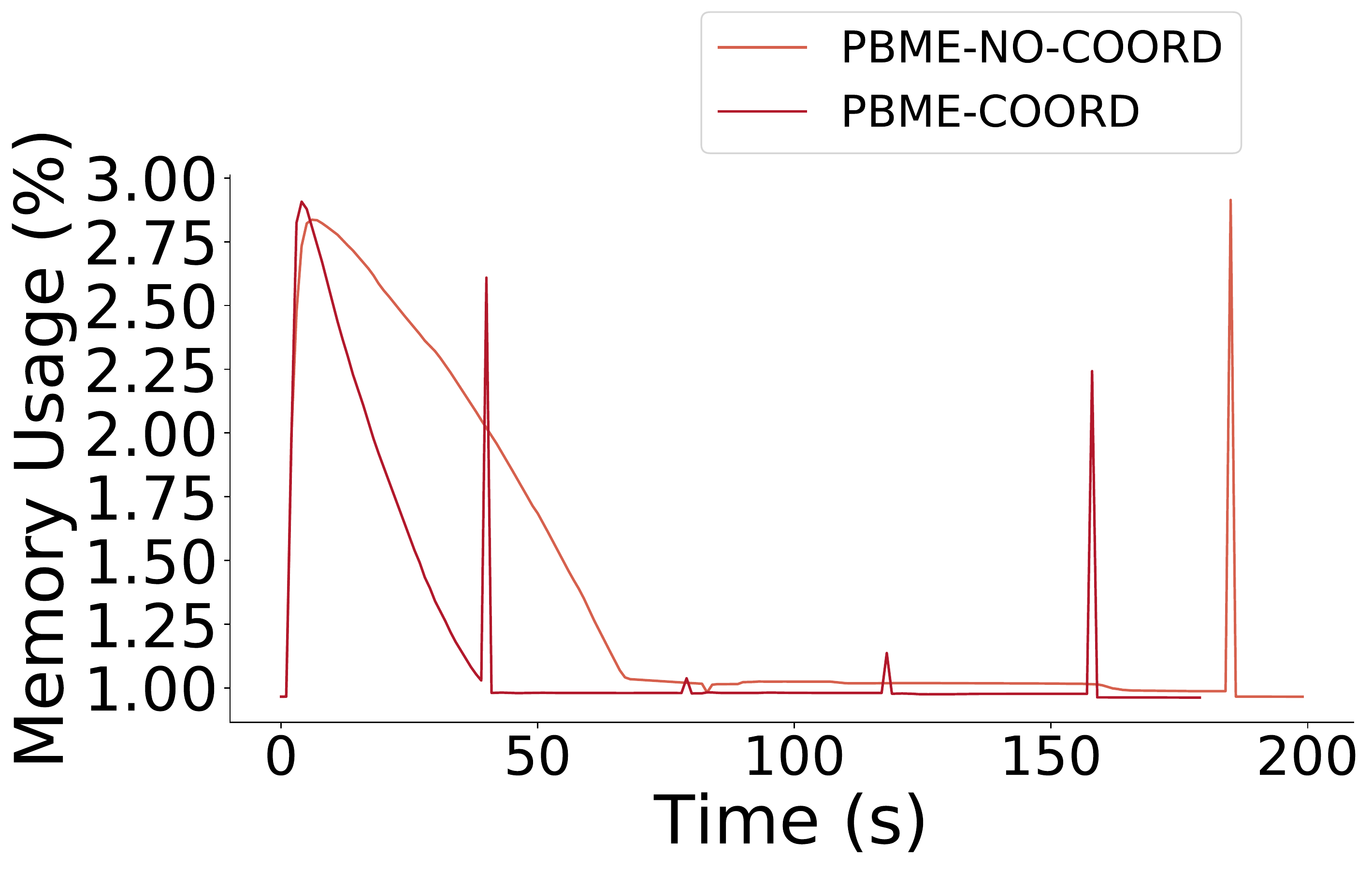}
        \caption{Memory Usage}
        \label{fig:tc_alt_memory}
    \end{subfigure}
    \caption{({\tt G20K}) SG-PBME Coordination v.s Non-Coordination: As shown in Figure \ref{fig:sg_alt_memory}, with proper coordination, the CPU utilization of PBME with coordination achieves almost $100\%$ throughout the whole evaluation of SG on {\tt G20k} and it takes less time to finish compared to PBME without coordination. There is no much difference in memory consumption of two methods as shown in Figure \ref{fig:tc_alt_memory}.}
    \label{fig:sg_alt_memory_cpu}
\end{figure}

The \textsc{pbme} of SG is outlined in Algorithm~\ref{alg:bit_matrix_eval_sg}, and has some differences compared
to that of TC. 
In addition to the output matrix, an additional index is built on the $\edb$ relation ${\tt arc{(x,y)}}$ (line~\ref{alg:bit_matrix_eval_sg:index}), since the $\edb$ relation ${\tt arc{(x,y)}}$ appears twice in the recursive rule, direct scanning the $\edb$ tables without using index (line~\ref{alg:bit_matrix_eval_sg:use_index1}-\ref{alg:bit_matrix_eval_sg:use_index2} uses index scan) might introduce great amount of redundant computation.
Unlike evaluation of TC, we note that issues such as data skew and redundant computation for SG are possibly seen in evaluation of different threads in different data partitions mainly because $\delta$ (Algorithm \ref{alg:bit_matrix_eval_sg} line \ref{alg:bit_matrix_eval_sg:delta} -\ref{alg:bit_matrix_eval_sg:update}) in Algorithm \ref{alg:bit_matrix_eval_sg} produced is not \textit{tied} to data partitions assigned to threads (in TC, each row $i$ in partition $p$ \textit{always} updates the bit in row $i$, a.k.a $M_{tc}[i,]$ -   Algorithm \ref{alg:bit_matrix_eval_tc} line \ref{alg:bit_matrix_eval_tc:compute_delta}). To analyze the effect of data skew, we implement an experimental SG-PBME \textit{with coordination} (SG-PBME-COORD) and compare it with SG-PBME \textit{without coordination} as shown in Figure \ref{fig:sg_alt_memory_cpu}. SG-PBME-COORD mitigates the data skew by \textit{aggregating} and \textit{re-balancing} - when a thread has generated $\delta$ the size of which is above a given threshold $t$, the $\delta$ is aggregated and packed as a \textit{work order}, being sent to the \textit{global pool}. Threads that are idle will grab work orders from the global pool to achieve load balancing. There is a trade-off incurred by the choice of $t$ - if $t$ is too small, then there will be too much communication overheads incurred by message passing, slowing down the performance; on the contracy, if $t$ is too large, then workload balancing cannot be well achieved when the skew is serious. Meanwhile, coordination will only incur unnecessary overhead when there is nearly no skew in data. We leave all the detailed analysis and study of this coordination approach in our future work.

\section{Experimental Evaluation}
\label{sec:evaluation}

In this section, we evaluate the performance of \datalogsys. Our experimental
evaluation focuses on answering the following two questions:
\begin{packed_enum}
\item How does our proposed system scale with increased computation power (cores) and data size?
\item How does our proposed system perform compared to other parallel \datalog\ evaluation engines?
\end{packed_enum}
To answer these two questions, we perform experiments using several benchmark \datalog\ programs from the literature: both from
traditional graph analytics tasks (e.g., reachability, shortest path, connected components),
as well as program analysis tasks (e.g., pointer static analysis).
We compare \datalogsys\ against a variety of state-of-the-art \datalog\ engines, as well as a recent single-machine, multicore engine (Graspan), which can express only a subset of \datalog.

\subsection{Experimental Setup}
We briefly describe here the setup for our experiments.

\introparagraph{System Configuration.}
Our experiments are conducted on a bare-metal server in Cloudlab~\cite{cloudlab}, a large cloud infrastructure.
The server runs Ubuntu 14.04 LTS and has two Intel Xeon E5-2660 v3 2.60 GHz (Haswell EP) processors.
Each processor has 10 cores, and 20 hyper-threading hardware threads.
The server has 160GB memory and each NUMA
node is directly attached to 80GB of memory.

\introparagraph{Other Datalog Engines.}
We compare the performance of \datalogsys\ with several state-of-the-art systems that perform either
general \datalog\ evaluation, or evaluate only a fragment of \datalog\ for domain-specific tasks.


\begin{packed_enum}
\item  BigDatalog~\cite{BigDatalog} is a general-purpose distributed \datalog\ system implemented on top of Spark.\footnote{
BigDatalog exhibits significant performance improvements over Myria
and Socialite, and therefore  we do not compare against them.}

\item Souffle~\cite{Souffle} is a parallel \datalog\ evaluation tool that compiles \datalog\ to a native C++ program. It focuses on evaluating \datalog\ programs for the domain of static program analysis.\footnote{
Recent work has shown that Souffle outperforms LogicBlox~\cite{antoniadis2017porting}.
Indeed, our early attempts using LogicBlox confirm that its performance is not
comparable to other parallel \datalog\ systems. Thus, we exclude LogixBlox from our experimental evaluation.}

\item \bddbddb~\cite{bddbddb} is a single-thread \datalog\ solver designed for static program analysis. Its key feature is the representation of relations using binary decision diagrams (\bdds).
\item Graspan~\cite{Graspan} is a single-machine disk-based parallel graph system, used mainly for interprocedural static analysis of large-scale system code.
\end{packed_enum}

\begin{table*}
\begin{center}
\begin{tabular} { c c c }
\toprule
\textbf{Graph Analytics} &\ \textbf{Datasets} &\ \textbf{Reference} \\
\midrule
Transitive Closure (TC) &\ [{\tt G5K}, {\tt G10K}, {\tt G10K-0.01}, {\tt G10K-0.1}, {\tt G20K}, {\tt G40K}, {\tt G80K}] &\ \cite{BigDatalog} \\

Same Generation (SG) &\ [{\tt G5K}, {\tt G10K}, {\tt G10K-0.01}, {\tt G10K-0.1}, {\tt G20K}, {\tt G40K}, {\tt G80K}]  &\ \cite{BigDatalog} \\

Reachability (REACH) &\ [{\tt livejournal}, {\tt orkut}, {\tt arabic}, {\tt twitter}], {\tt RMAT} &\ \cite{BigDatalog}\\

Connected Components (CC) &\ [{\tt livejournal}, {\tt orkut}, {\tt arabic}, {\tt twitter}], {\tt RMAT} &\ \cite{BigDatalog} \\

Single Source Shortest Path (SSSP) &\ [{\tt livejournal}, {\tt orkut}, {\tt arabic}, {\tt twitter}], {\tt RMAT}&\ \cite{BigDatalog}\vspace{1ex} \\

\toprule
\textbf{Program Analysis} &\ \textbf{Datasets}  &\ \textbf{Reference} \\
\midrule

Andersen's Analysis (AA) &\ 7 synthetic datasets &\ - \\

Context-sensitive Dataflow Analysis (CSDA)  &\ [{\tt linux}, {\tt postgresql}, {\tt httpd}] &\ \cite{Graspan} \\

Context-sensitive Points-to Analysis (CSPA) &\ [{\tt linux}, {\tt postgresql}, {\tt httpd}] &\ \cite{Graspan} \\
\bottomrule
\end{tabular}
\caption{Summary of Datalog Programs and Datasets in Performance Evaluation}
\label{tab:benchmark}
\end{center}
\end{table*}

\subsection{Benchmark Programs and Datasets}

We conduct our experiments using \datalog\ programs that arise from two different domains: {\em graph analytics}
and {\em static program analysis}.
The graph analytics benchmarks are those used for evaluating
BigDatalog~\cite{BigDatalog}. Below, we present them in detail (with the exception of TC and SG, which
are described earlier in the paper).

\vspace{1ex}
\noindent
\textbf{\textit{Reachability}}
\begin{align*}
{\tt reach(y)} & \cdash {\tt id(y)}. \\
{\tt reach(y)} & \cdash {\tt reach(x), arc(x,y)}.
\end{align*}
%

\noindent
\textbf{\textit{Connected Components}}
\begin{align*}
{\tt cc3(x, MIN(x))} & \cdash {\tt arc(x, \_)}.\\
{\tt cc3(y, MIN(z))} & \cdash {\tt cc3(x,z), arc(x,y)}.\\
{\tt cc2(x, MIN(y))} & \cdash {\tt cc3(x,y)}. \\
{\tt cc(x)} & \cdash  {\tt cc2(\_,x)}.
\end{align*}

\noindent
\textbf{\textit{Single Source Shortest Path}}
\begin{align*}
{\tt sssp2(y, MIN(0))} & \cdash {\tt id(y)}.\\
{\tt sssp2(y, MIN(d1+d2))} & \cdash  {\tt sssp2(x, d1), arc(x, y, d2)}.\\
{\tt sssp(x, MIN(d))} & \cdash {\tt sssp2(x,d)}.
\end{align*}

The static analysis benchmarks include analyses on which Graspan was evaluated~\cite{Graspan}, as well as a classic static analysis called Andersen's analysis~\cite{andersen1994program}.

\vspace{1ex}
\noindent
\textbf{\textit{Andersen's Analysis}}
\begin{align*}
{\tt pointsTo(y,x)} & \cdash  {\tt addressOf(y,x)}.\\
{\tt pointsTo(y,x)} & \cdash  {\tt assign(y,z), pointsTo(z,x)}.\\
{\tt pointsTo(y,w)} & \cdash  {\tt load(y,x), pointsTo(x,z), pointsTo(z,w)}.\\
{\tt pointsTo(z,w)} & \cdash  {\tt store(y,x), pointsTo(y,z), pointsTo(x,w)}.
\end{align*}

\noindent
\textbf{\textit{Context-sensitive Points-to Analysis} (CSPA)}\footnote{
Graspan's analysis is context-sensitive via method cloning~\cite{whaley04}---therefore, calling context does not appear
in the rules, but in the data.}
\begin{align*}
{\tt valueFlow(y,x)} & \cdash {\tt assign(y,x)}.\\
{\tt valueFlow(x,y)} & \cdash {\tt assign(x,z), memoryAlias(z,y)}.\\
{\tt valueFlow(x,y)} & \cdash {\tt valueFlow(x,z), valueFlow(z,y)}.\\
{\tt memoryAlias(x,w)} & \cdash {\tt dereference(y,x), valueAlias(y,z)}, \\
					      &   {\tt dereference(z,w)}.\\
{\tt valueAlias(x,y)} & \cdash {\tt valueFlow(z,x), valueFlow(z,y)}.\\
{\tt valueAlias(x,y)} & \cdash {\tt valueFlow(z,x), memoryAlias(z,w)}, \\
			                   & {\tt valueFlow(w,y)}.\\
{\tt valueFlow(x,x)} & \cdash {\tt assign(x,y)}.\\
{\tt valueFlow(x,x)} & \cdash {\tt assign(y,x)}.\\
{\tt memoryAlias(x,x)} & \cdash {\tt assign(y,x)}.\\
{\tt memoryAlias(x,x)} & \cdash {\tt assign(x,y)}.
\end{align*}

\noindent
\textbf{\textit{Context-sensitive Dataflow Analysis} (CSDA)}\\
(\emph{uses results of CSPA})
\begin{align*}
{\tt null(x,y)} & \cdash {\tt nullEdge(x,y)}.\\
{\tt null(x,y)} & \cdash {\tt null(x,w)}, {\tt arc(w,y)}.
\end{align*}

To evaluate the benchmark programs, we use a combination of synthetic and real-world datasets,
which are summarized in Table~\ref{tab:benchmark}. To give a better view of the performance evaluation, 
we briefly summarize some of the datasets and corresponding Datalog programs here. For more details, readers can go to 
the reference. 

G$n$-$p$ graphs are graphs generated by the GTgraph synthetic graph generator, where $n$ represents the 
number of total vertices of the graph in which each pair of vertices is connected by probability $p$. Each pair of vertices in 
G$n$ omitting $p$ is connected with probability $0.001$. All G$n$-$p$ graphs are very dense considering their relatively small 
number of vertices. SG and TC generate very large results when evaluation is performed on G$n$-$p$,  (at least a few orders of 
magnitude larger than the number of vertices). {\tt RMAT} graphs are graphs generated by the RMAT graph generator, 
with the same specification in \cite{BigDatalog}, {\tt RMAT-n} represents the graph that has $n$ vertices and 10$n$ directed edges 
($n \in {1M, 2M, 4M, 8M, 16M, 32M, 64M, 128M}$ in our evaluation experiments). {\tt livejournal}, {\tt orkut}, {\tt arabic}, {\tt twitter} 
are all large-scale real-world graphs which have tens of millions of vertices and edges. For the Andersen analysis, seven datasets
are generated ranging from small size to large size based on the characteristics of a tiny real dataset available at hand, numbered from 1 to 7. 
The graph representations of the datasets are small and produce moderate number of tuples. {\tt linux}, {\tt postgresql}, {\tt httpd} are all real 
system programs used for CSDA and CSPA experiments in \cite{Graspan}.

\subsection{Experimental Results}
We first evaluate  the scalability of our proposed engine, and then move
to a comparison with other systems.


\begin{figure}[t]
\centering
    \begin{subfigure}[b]{0.23\textwidth}
        \centering
        \includegraphics[scale=0.11]{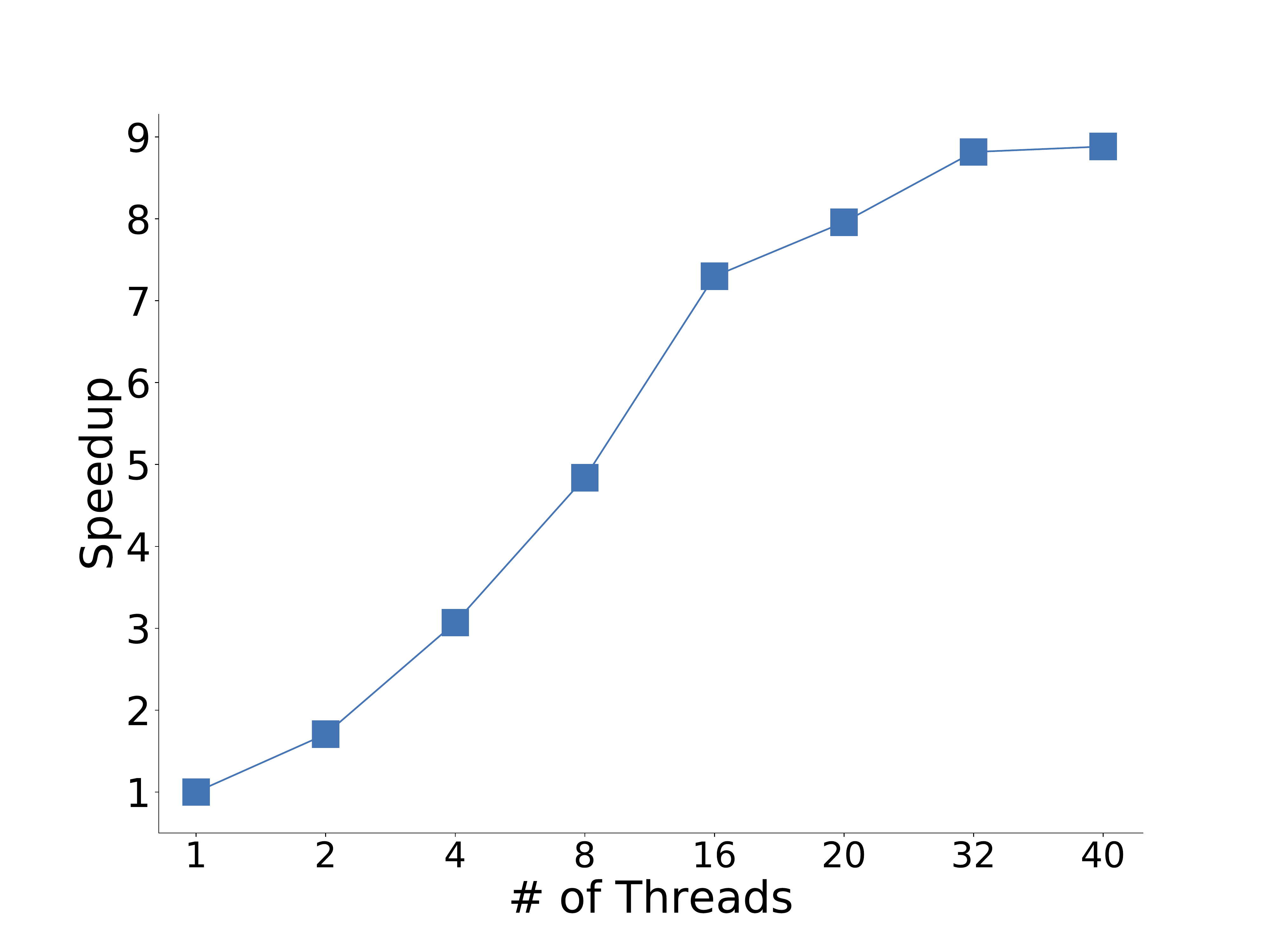}
        \caption{CSPA on {\tt httpd} }
        \label{fig:quickstep_speedup_cc}
    \end{subfigure}
    \begin{subfigure}[b]{0.23\textwidth}
        \centering
        \includegraphics[scale=0.11]{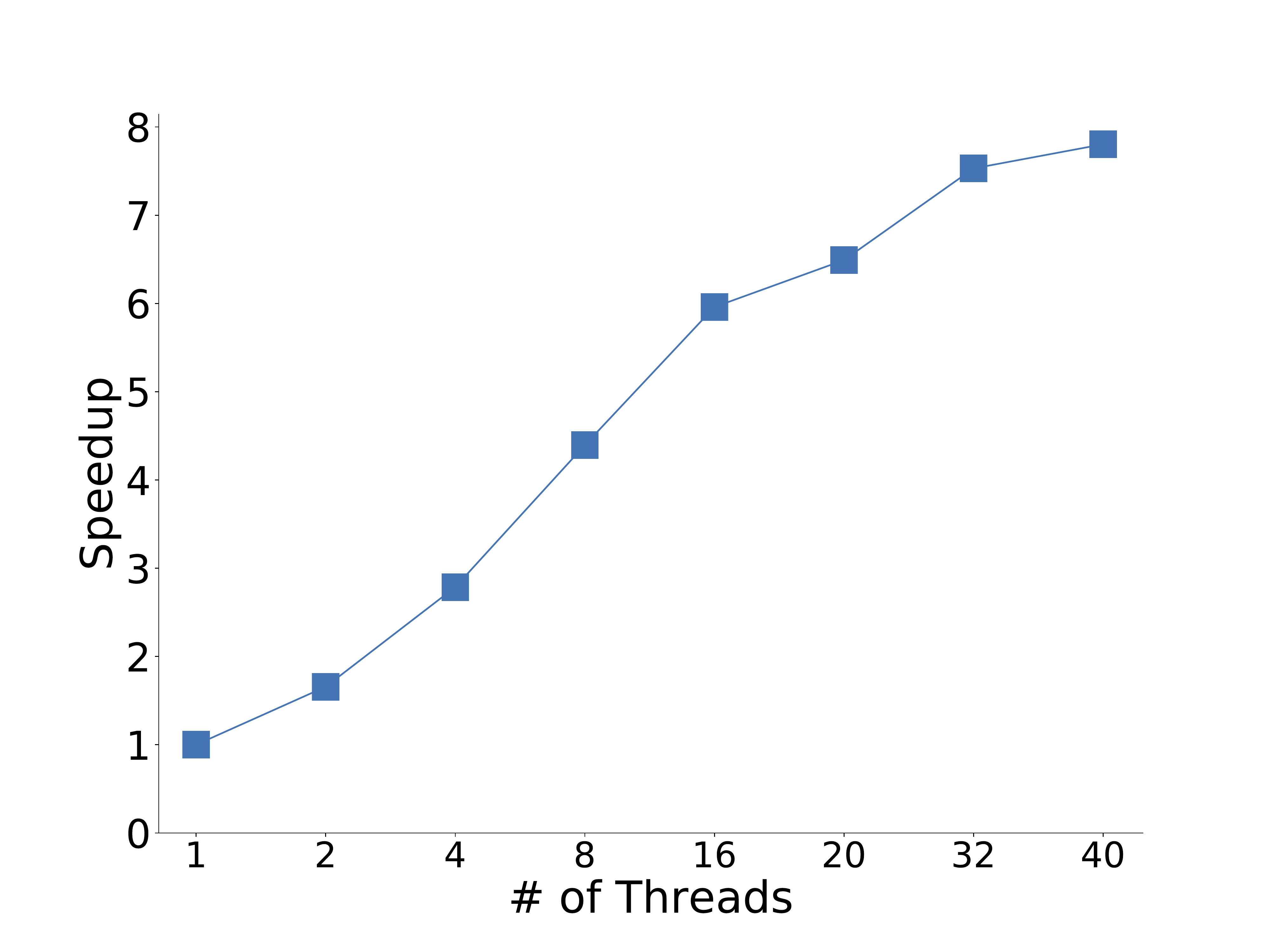}
        \caption{CC on {\tt livejournal}}
        \label{fig:quickstep_speedup_cc}
    \end{subfigure}
    \caption{Scaling-up on Cores}
    \label{fig:scaling_up_on_cores}
\end{figure}


\introparagraph{Scaling-up Cores.} 
To evaluate the speedup of \datalogsys, we run the CC benchmark on {\tt livejournal} graph, and
the CSPA benchmark on the {\tt httpd} dataset. We vary the number of threads from 2 to 40. Figure \ref{fig:scaling_up_on_cores} demonstrates that for both cases, \datalogsys\ scales really well using up to 16 threads, and after that point achieves a much smaller speedup. This drop on speedup occurs because of the synchronization/scheduling primitive around the common shared hash table that is accessed from all the threads. Recall that 20 is the total number of physical cores.

\begin{figure}[H]
    \centering
    \begin{subfigure}[b]{0.23\textwidth}
        \centering
        \includegraphics[scale=0.11]{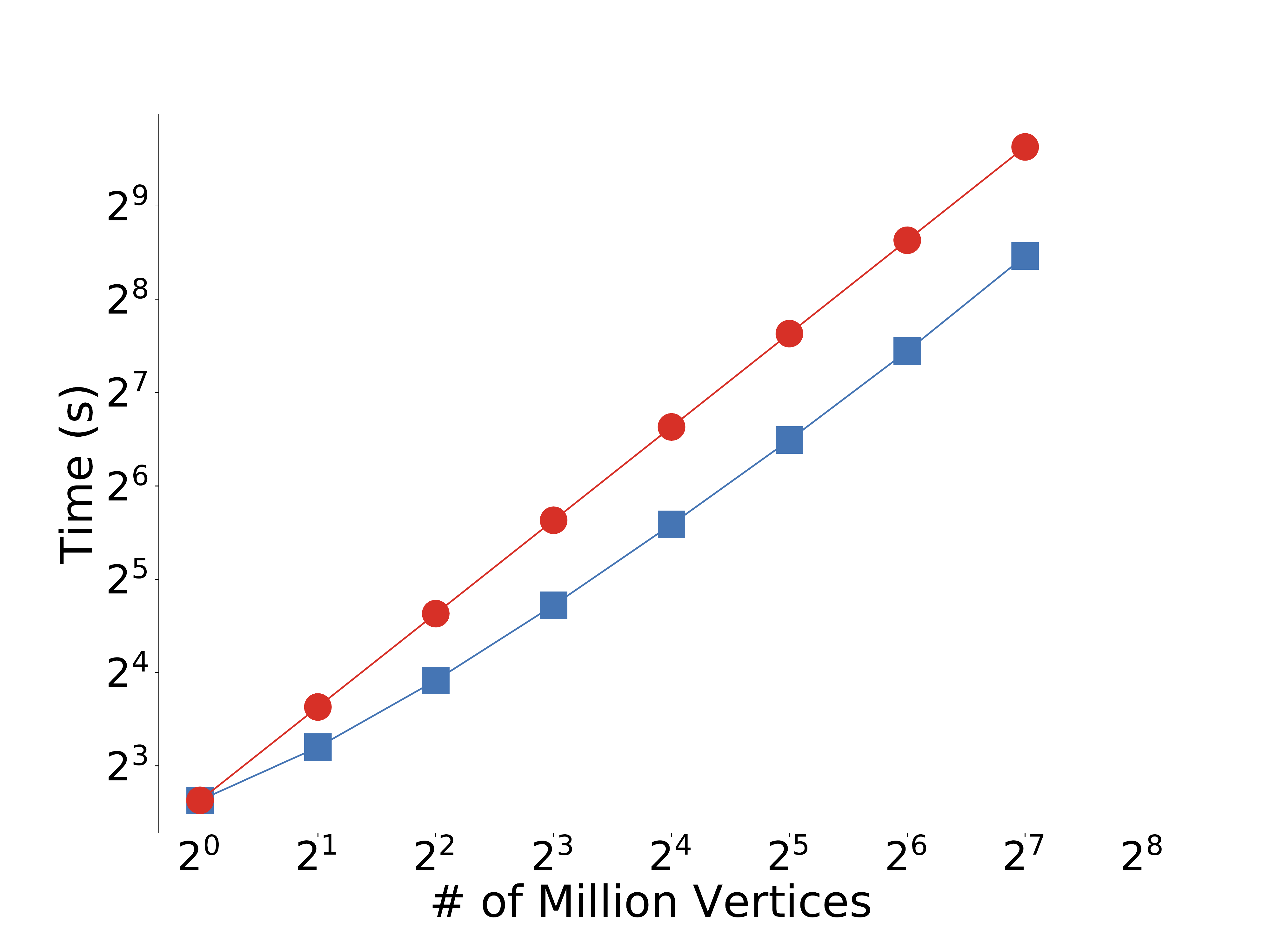}
        \caption{CC on {\tt RMAT}}
        \label{fig:relational_datalog_scale_rmat}
    \end{subfigure}
    \begin{subfigure}[b]{0.23\textwidth}
        \centering
        \includegraphics[scale=0.11]{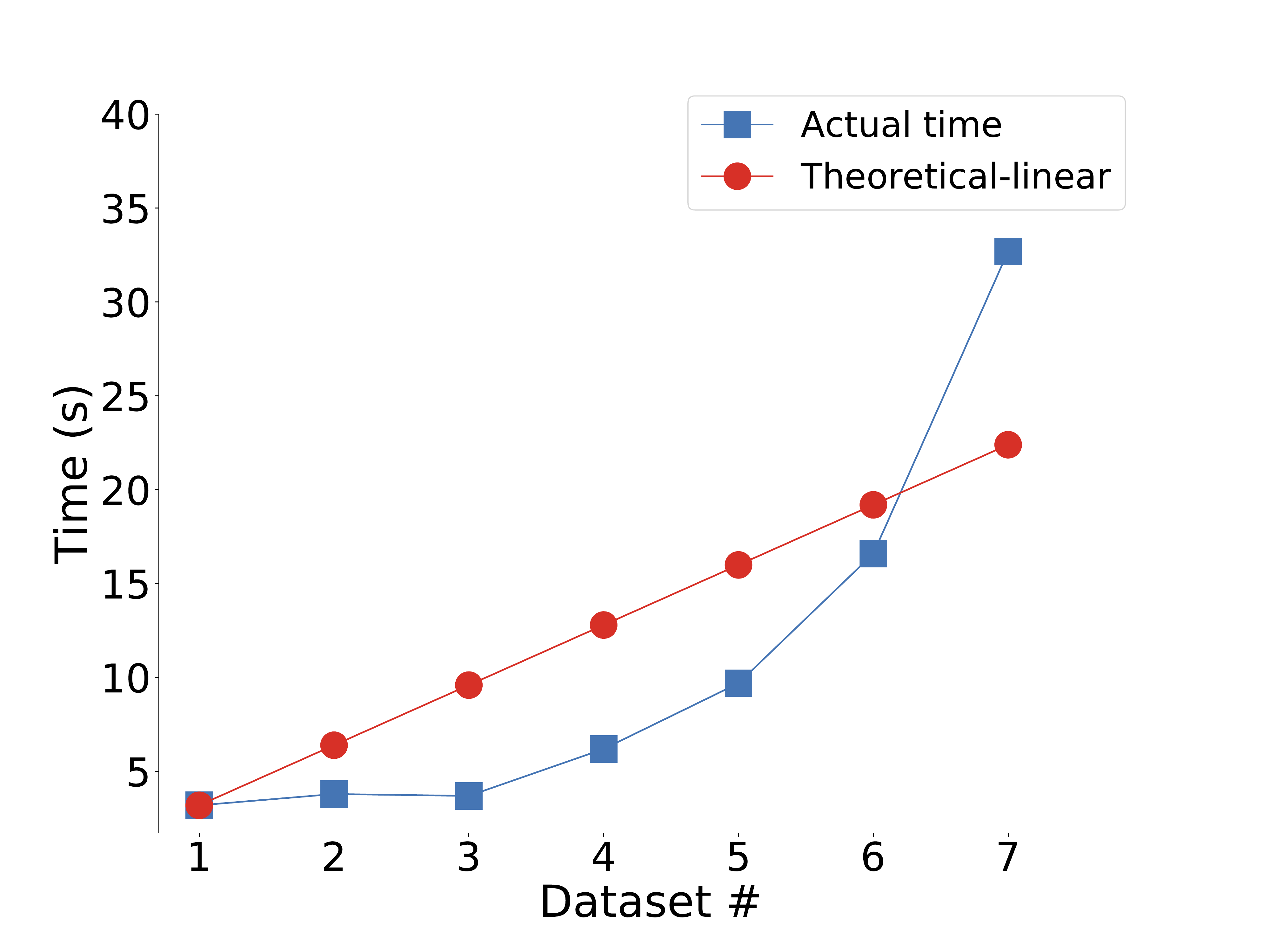}
        \caption{AA on synthetic dataset}
        \label{fig:relational_datalog_scale_andersen}
    \end{subfigure}
    \caption{Scaling-up on Datasets: the x-axis of \ref{fig:relational_datalog_scale_rmat} represents the number of vertices of the corresponding {\tt RMAT} graph datasets ({\tt RMAT-1M} to {\tt RMAT-128M});
    		 the synthetic datasets are numbered from 1 to 7 and the x-axis of \ref{fig:relational_datalog_scale_andersen} suggests the corresponding dataset number.
		 }
    \label{fig:scaling_up_on_datasets}
\end{figure}

\begin{figure}[t]
    \centering
    \begin{subfigure}[b]{0.5\textwidth}
        \centering
        \includegraphics[scale=0.22]{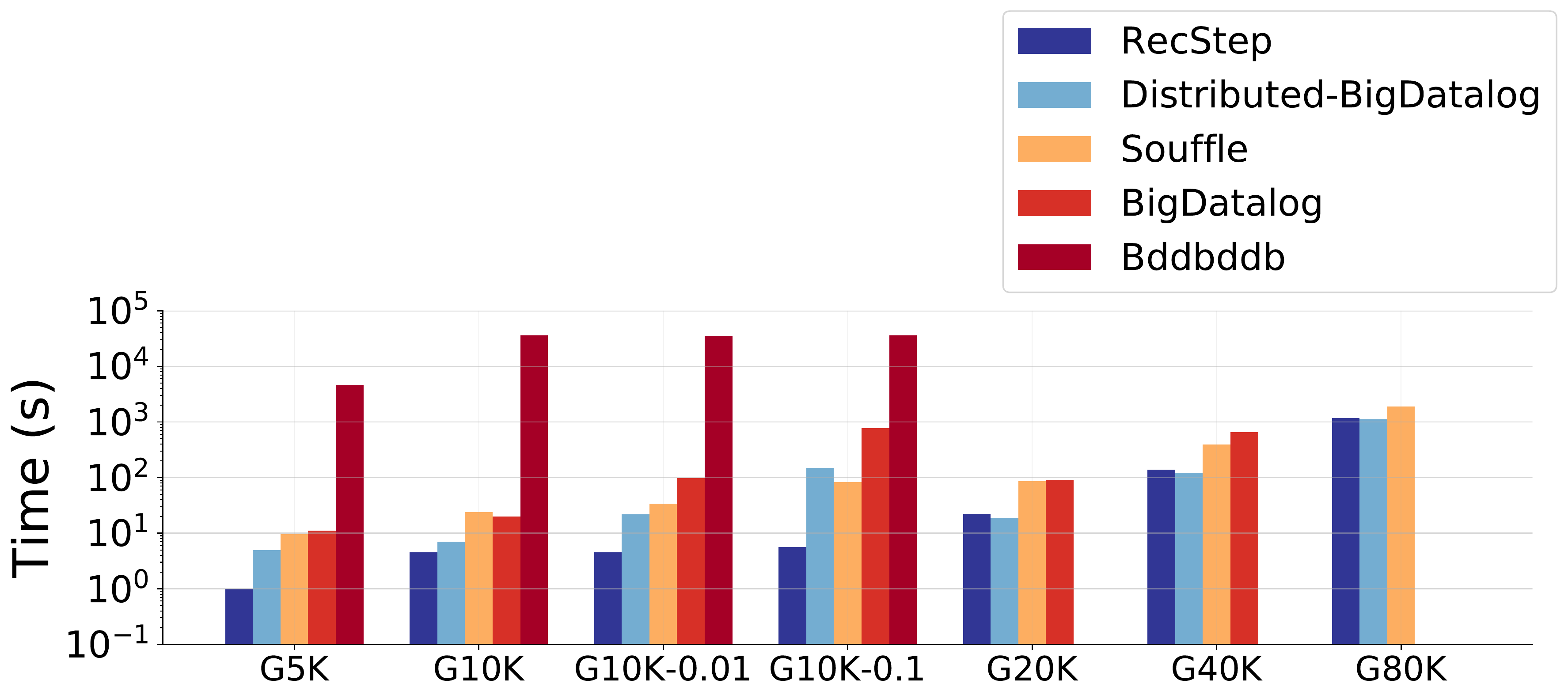}
        \caption{Transitive Closure}
        \label{fig:tc}
    \end{subfigure}
    \begin{subfigure}[b]{0.5\textwidth}
        \centering
        \includegraphics[scale=0.22]{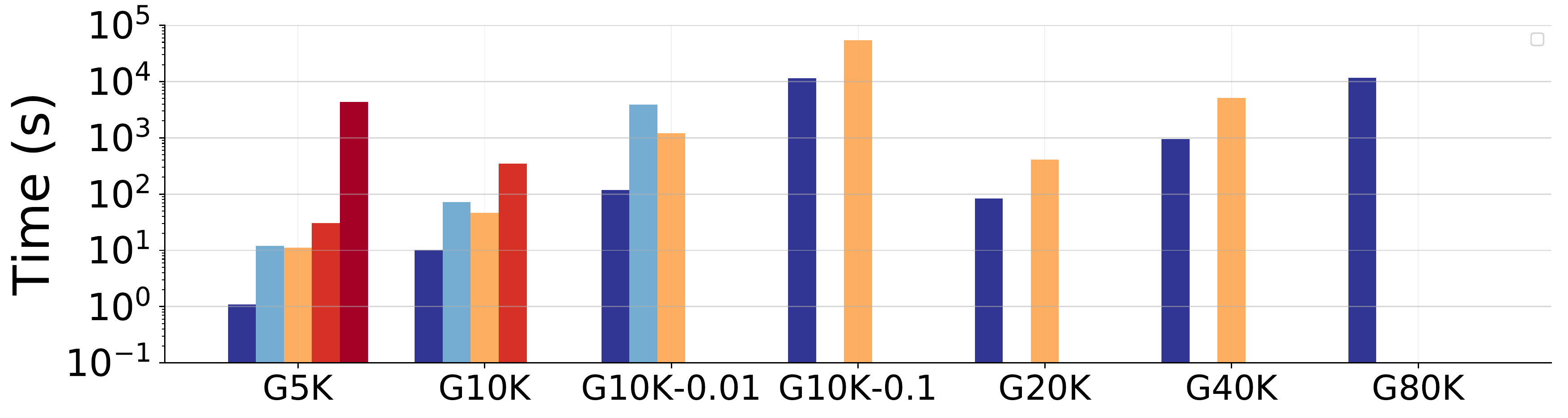}
        \caption{Same Generation}
        \label{fig:sg}
    \end{subfigure}
    \caption{Performance Comparison of TC and SG}
    \label{fig:tc_sg}
\end{figure}

\begin{figure}[t]
    \centering
    \begin{subfigure}[b]{0.23\textwidth}
        \centering
        \includegraphics[scale=0.15]{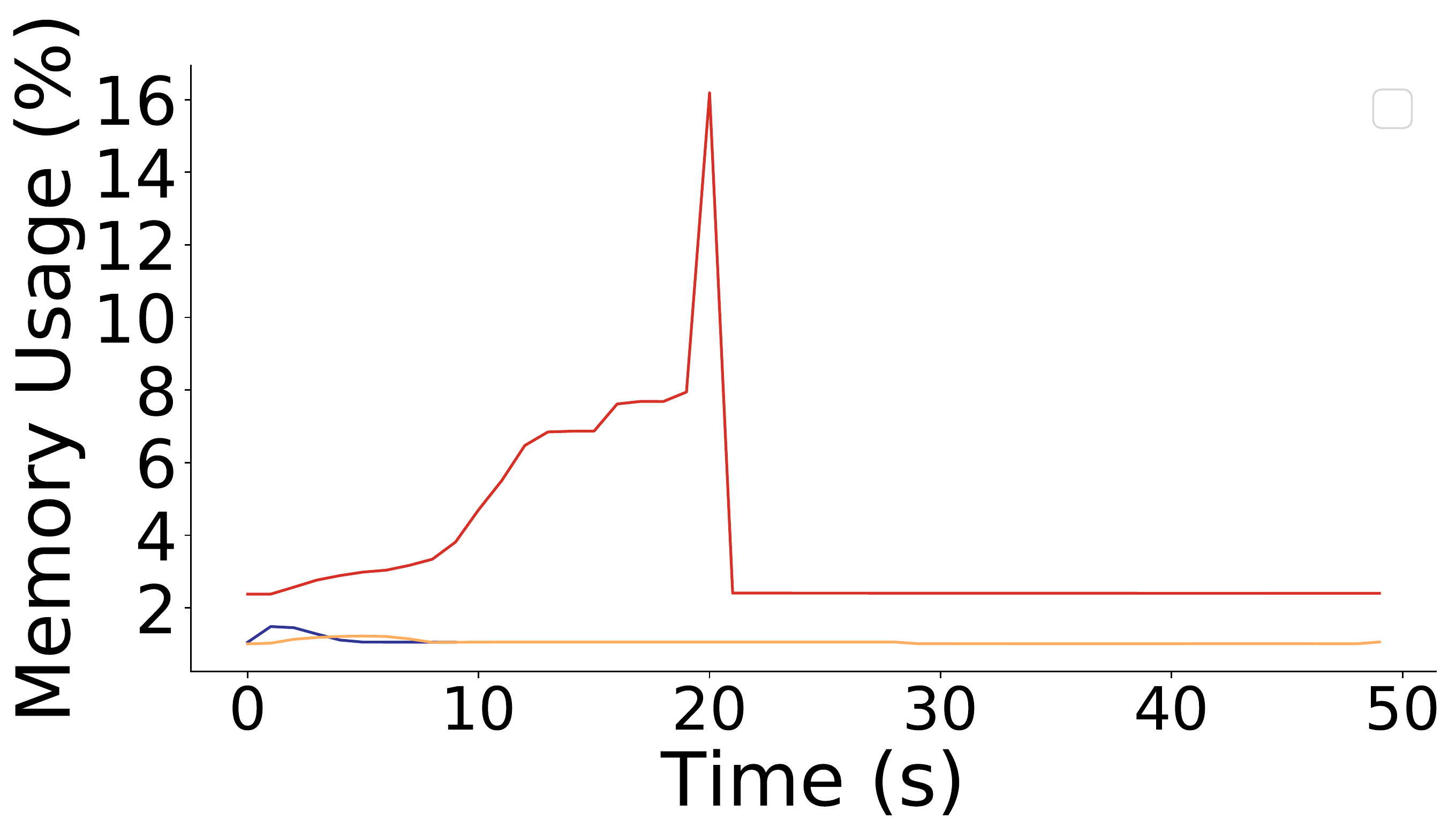}
        \caption{Transitive Closure}
        \label{fig:tc_memory}
    \end{subfigure}
    \begin{subfigure}[b]{0.23\textwidth}
        \centering
        \includegraphics[scale=0.15]{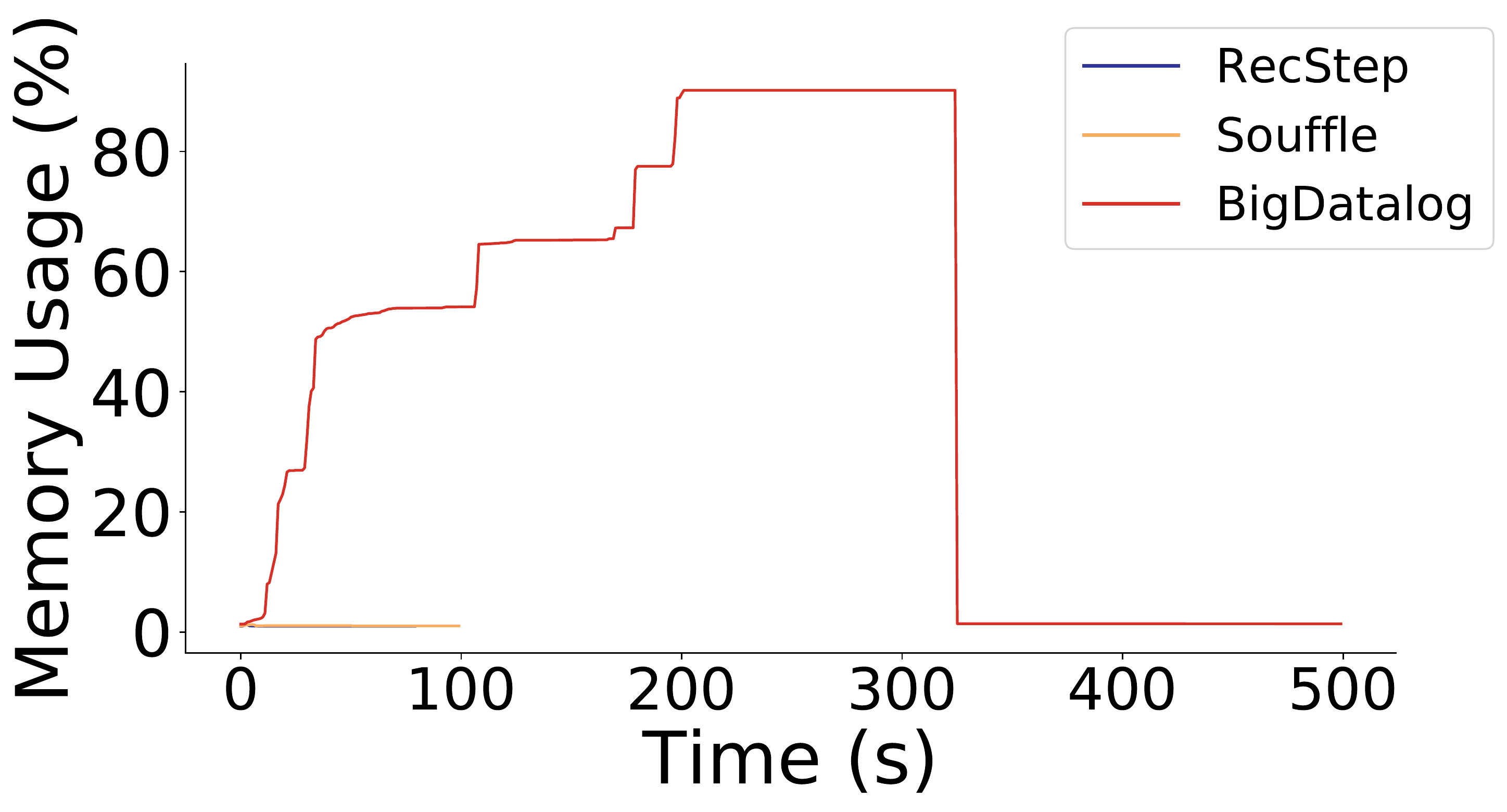}
        \caption{Same Generation}
        \label{fig:sg_memory}
    \end{subfigure}
    \caption{ Memory Usage of TC and SG (\tt{G10k})}
    \label{fig:tc_sg_memory}
\end{figure}

\begin{figure*}[t]
    \centering
    \begin{subfigure}[b]{0.3\textwidth}
        \centering
        \includegraphics[scale=0.18]{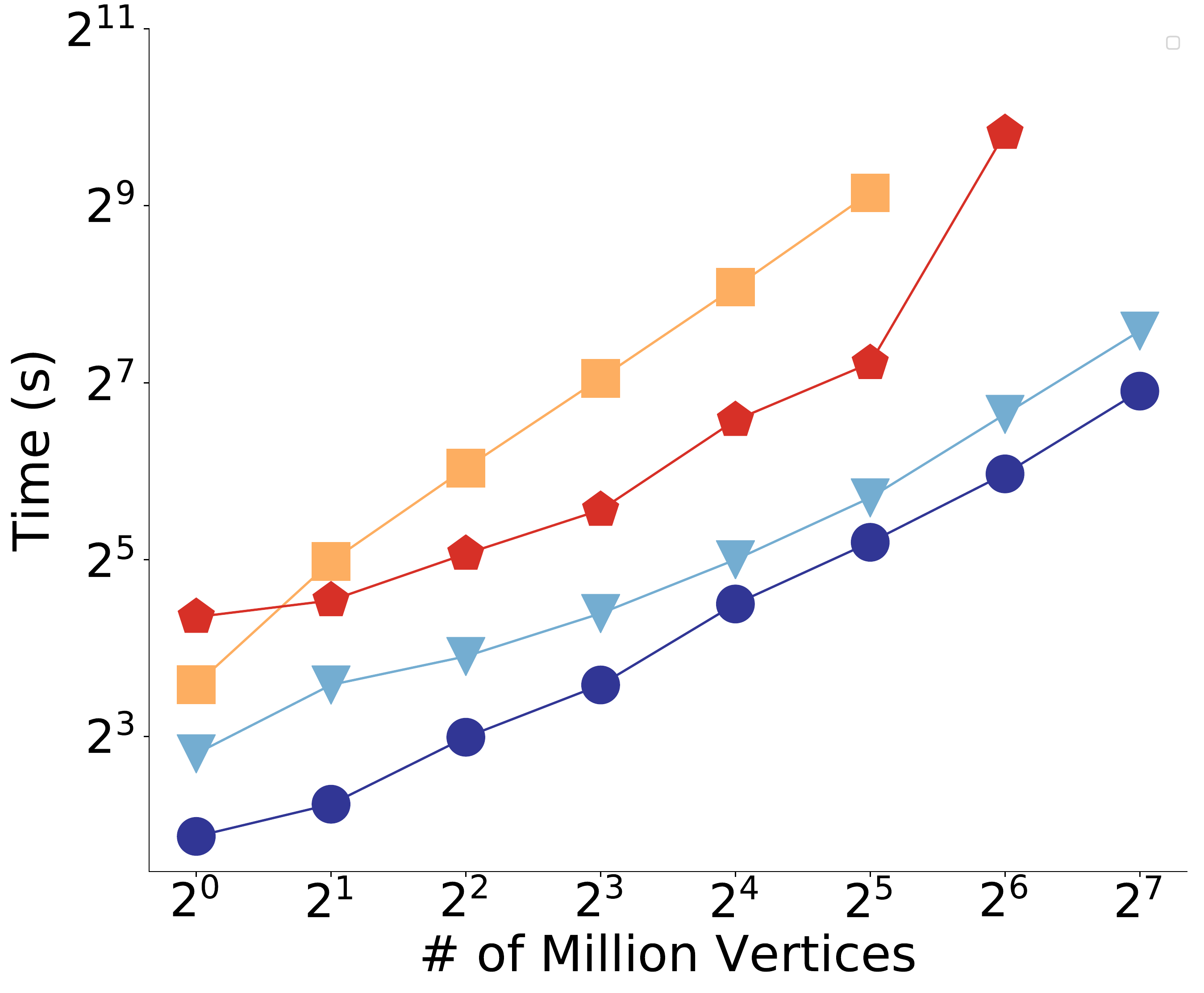}
        \caption{REACH}
        \label{fig:reach_rmat}
    \end{subfigure}
    \begin{subfigure}[b]{0.3\textwidth}
        \centering
        \includegraphics[scale=0.18]{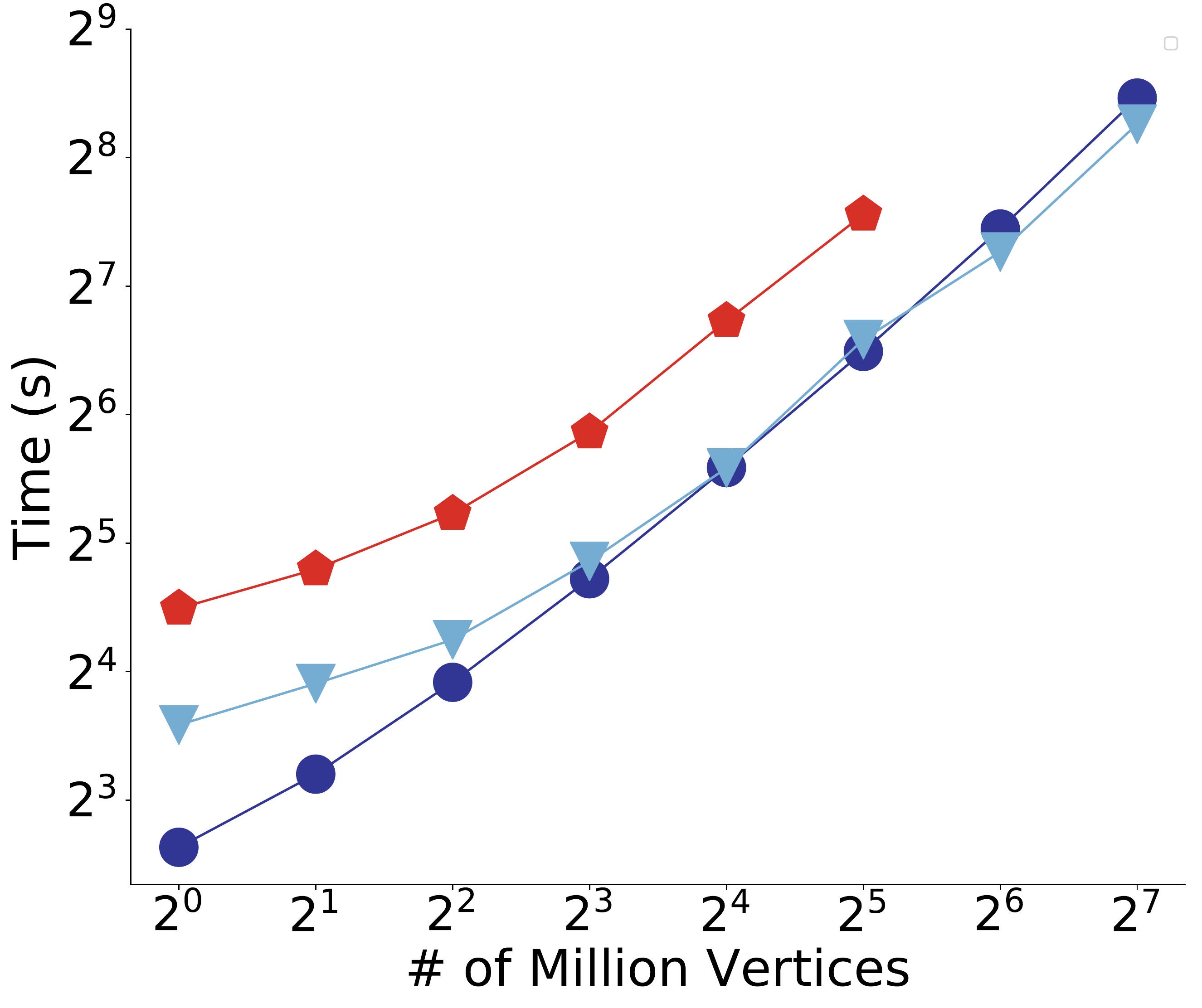}
        \caption{CC}
        \label{fig:cc_rmat}
    \end{subfigure}
    \begin{subfigure}[b]{0.3\textwidth}
        \centering
        \includegraphics[scale=0.18]{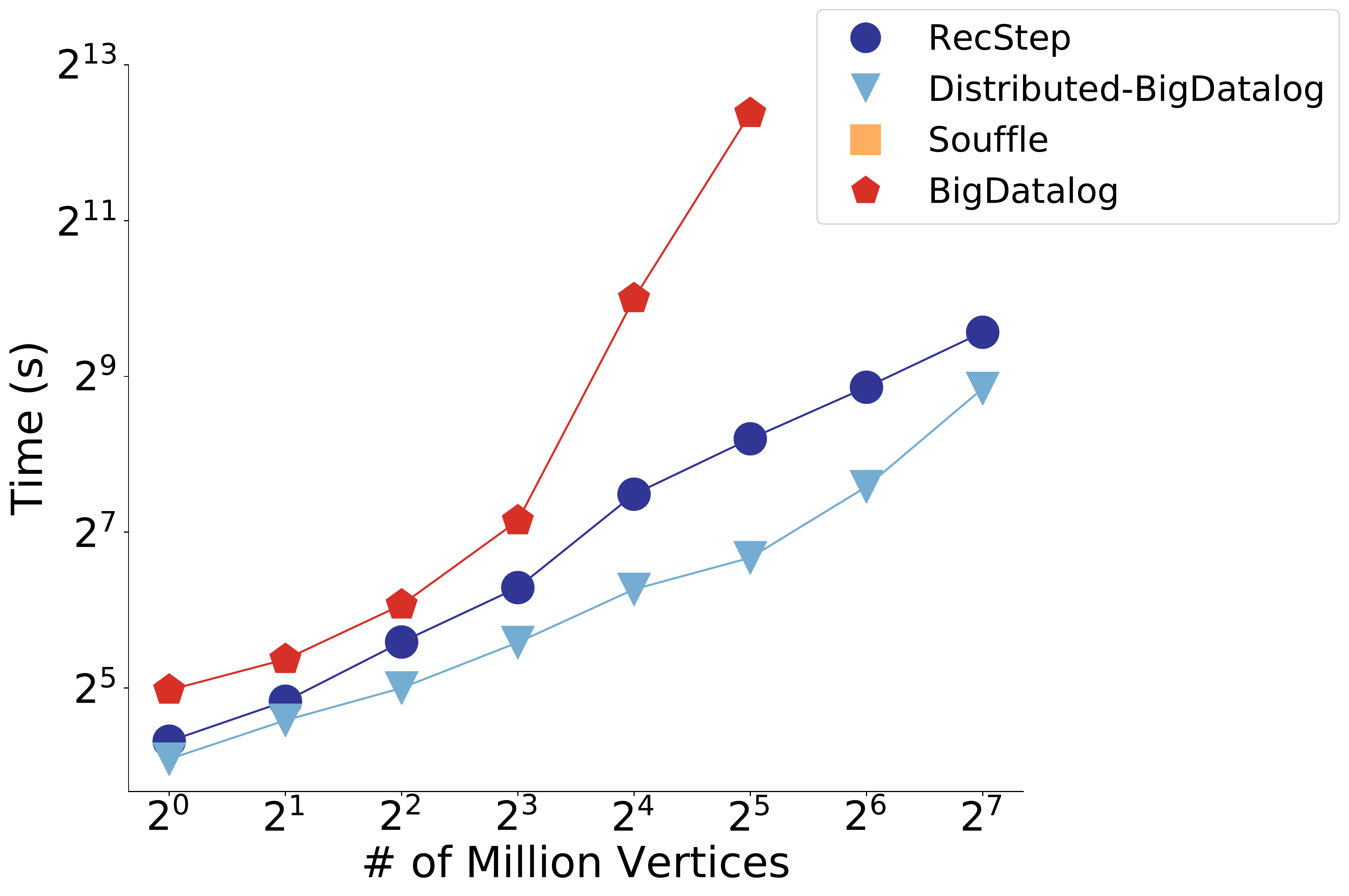}
        \caption{SSSP}
        \label{fig:sssp_rmat}
    \end{subfigure}
    \caption{Performance Comparison on {\tt RMAT} Graphs: The x-axis represents {\tt RMAT-1M} to {\tt RMAT-128M}.}
    \label{fig:rmat_exp}
\end{figure*}

\introparagraph{Scaling-up Data.}
We perform experiments on both a graph analytics program (CC on {\tt RMAT}) and a program analysis task (AA on the synthetic datasets) using all 20 physical cores (40 hyperthreads) to observe how our system scales over datasets of increasing sizes. From Figure \ref{fig:relational_datalog_scale_rmat}, we observe the time increases
nearly proportionally to the increasing size of the datasets.
 In Figure \ref{fig:relational_datalog_scale_andersen}, we observe that for datasets numbered from 1 to 3, the evaluation
times on these three datasets are roughly the same. 
The corresponding graphs representing the input for these three datasets are relatively sparse and
the total size of the data (input and intermediate results) during evaluation is small, and the cores/threads are underutilized; thus, 
when the data increases, the stale threads will take over the extra processing, and runtime will not increase. With the increasing size of datasets (4 to 7), 
we observe a similar trend as seen in Figure \ref{fig:relational_datalog_scale_rmat} since the size of input data as well as that of the produced intermediate results increases.
All cores are fully utilized, so more data will cause increase in runtime.

\begin{figure*}[t]
    \centering
    \begin{subfigure}[b]{0.3\textwidth}
        \centering
        \includegraphics[scale=0.18]{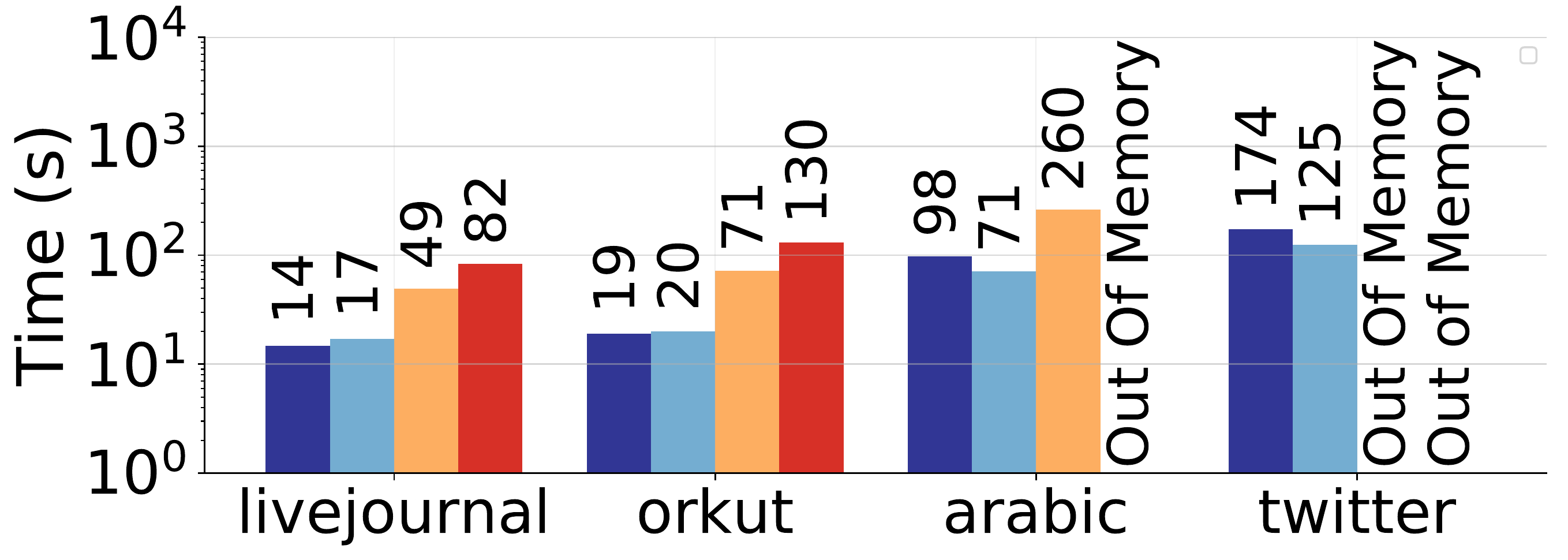}
        \caption{REACH}
        \label{fig:reach_real_world}
    \end{subfigure}
    \begin{subfigure}[b]{0.3\textwidth}
        \centering
        \includegraphics[scale=0.18]{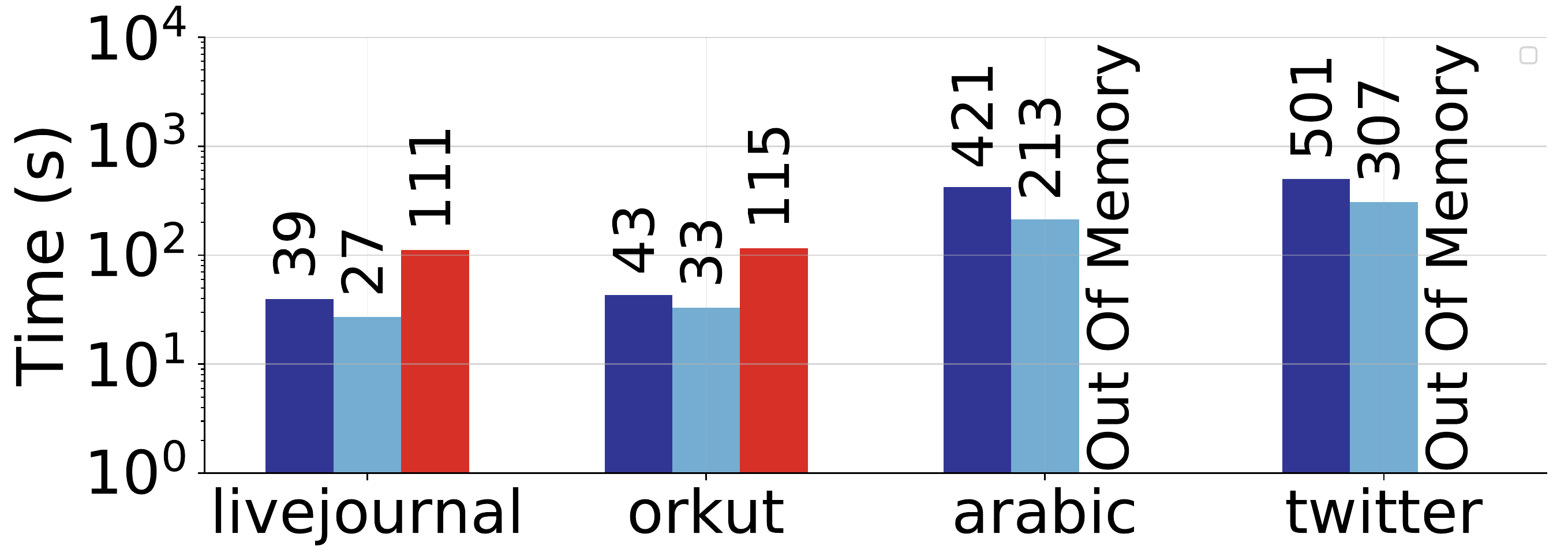}
        \caption{CC}
        \label{fig:cc_real_world}
    \end{subfigure}
    \begin{subfigure}[b]{0.3\textwidth}
        \centering
        \includegraphics[scale=0.18]{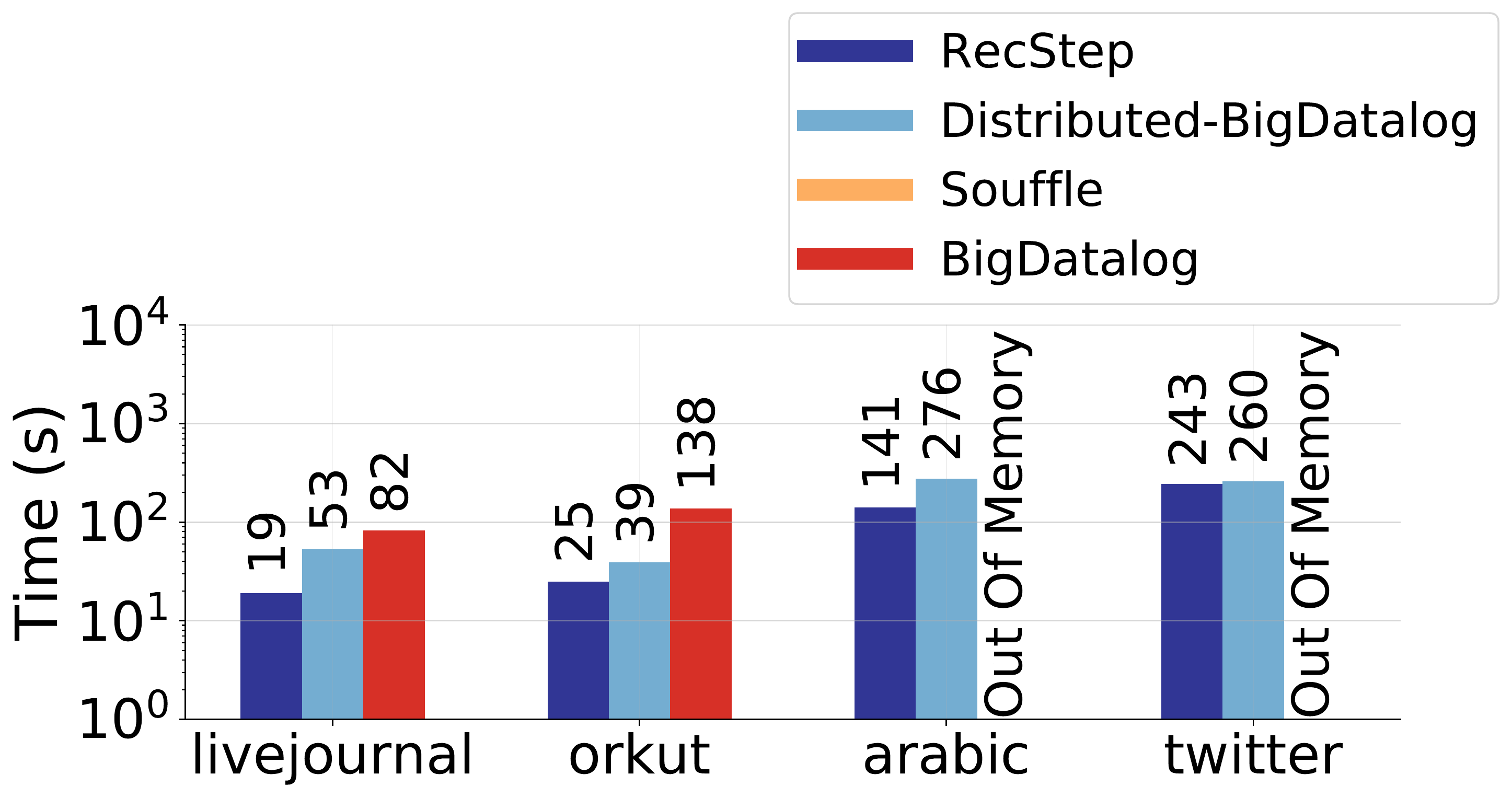}
        \caption{SSSP}
        \label{fig:sssp_real_world}
    \end{subfigure}
    \caption{Performance Comparison on Real-World Graphs.}
    \label{fig:real_world_exp}
\end{figure*}

\begin{figure*}[!ht]
    \centering
    \begin{subfigure}[b]{0.3\textwidth}
        \centering
        \includegraphics[scale=0.16]{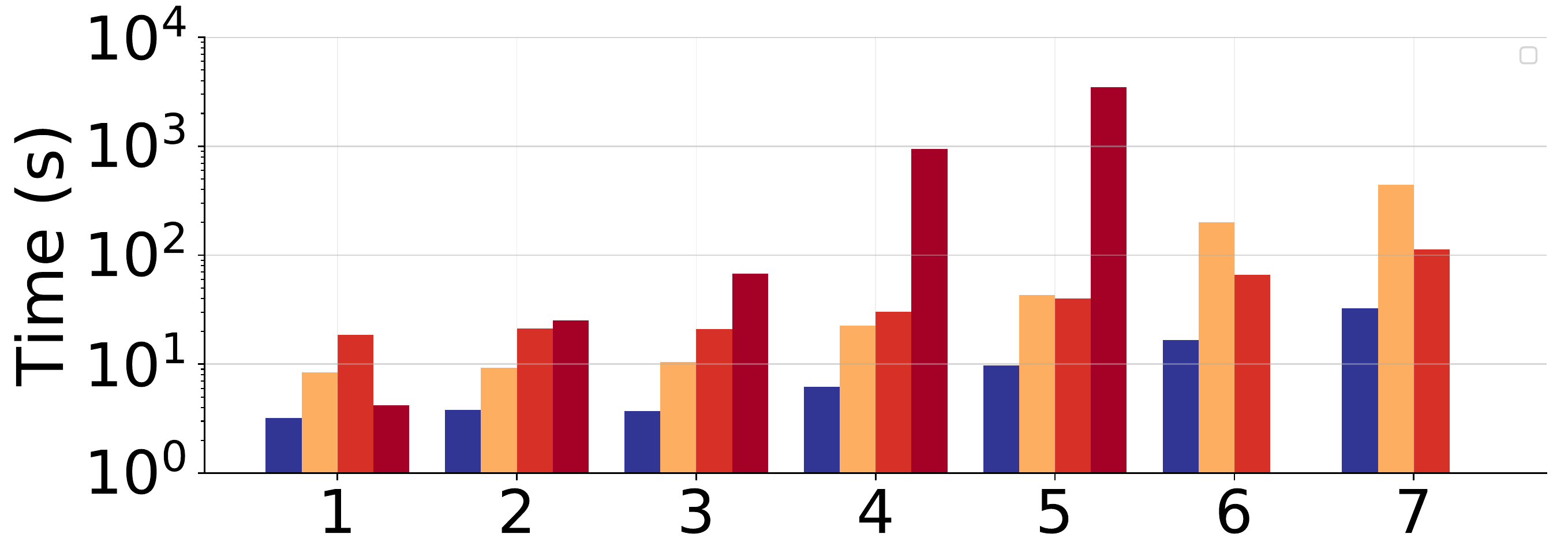}
        \caption{AA on 7 synthetic datasets}
        \label{fig:andersen_analysis}
    \end{subfigure}
    \begin{subfigure}[b]{0.3\textwidth}
        \centering
        \includegraphics[scale=0.16]{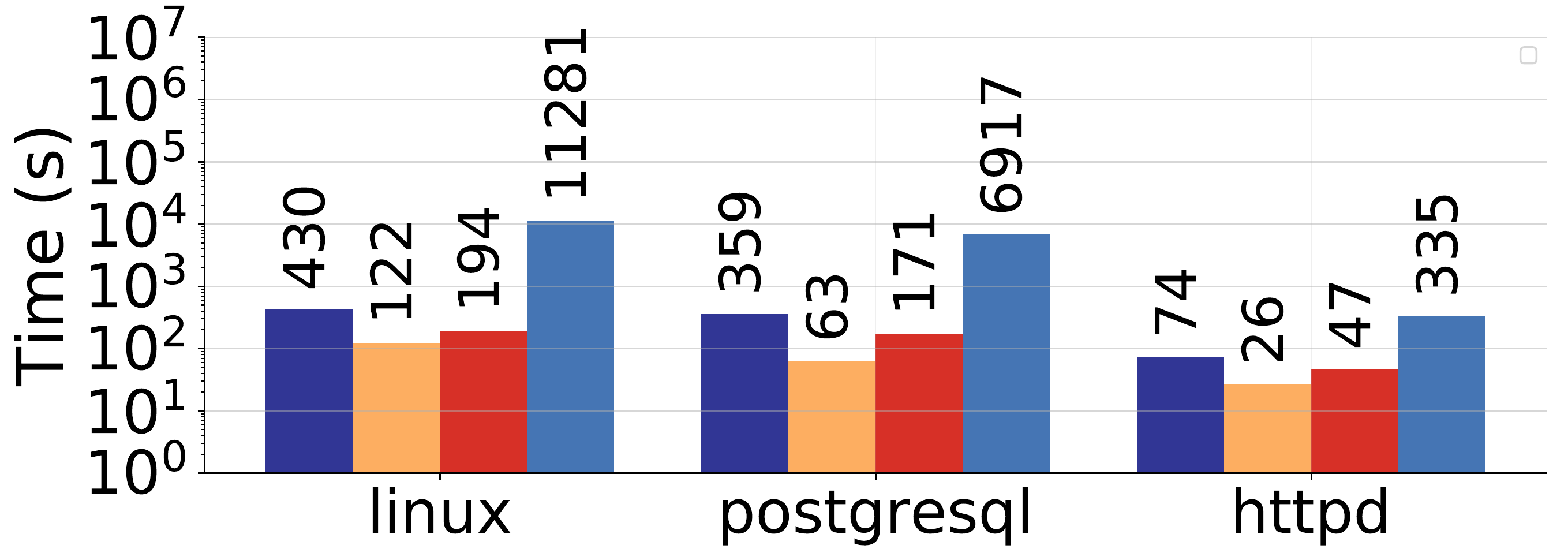}
        \caption{CSDA on systems programs}
        \label{fig:dataflow_analysis}
    \end{subfigure}
    \begin{subfigure}[b]{0.3\textwidth}
        \centering
        \includegraphics[scale=0.16]{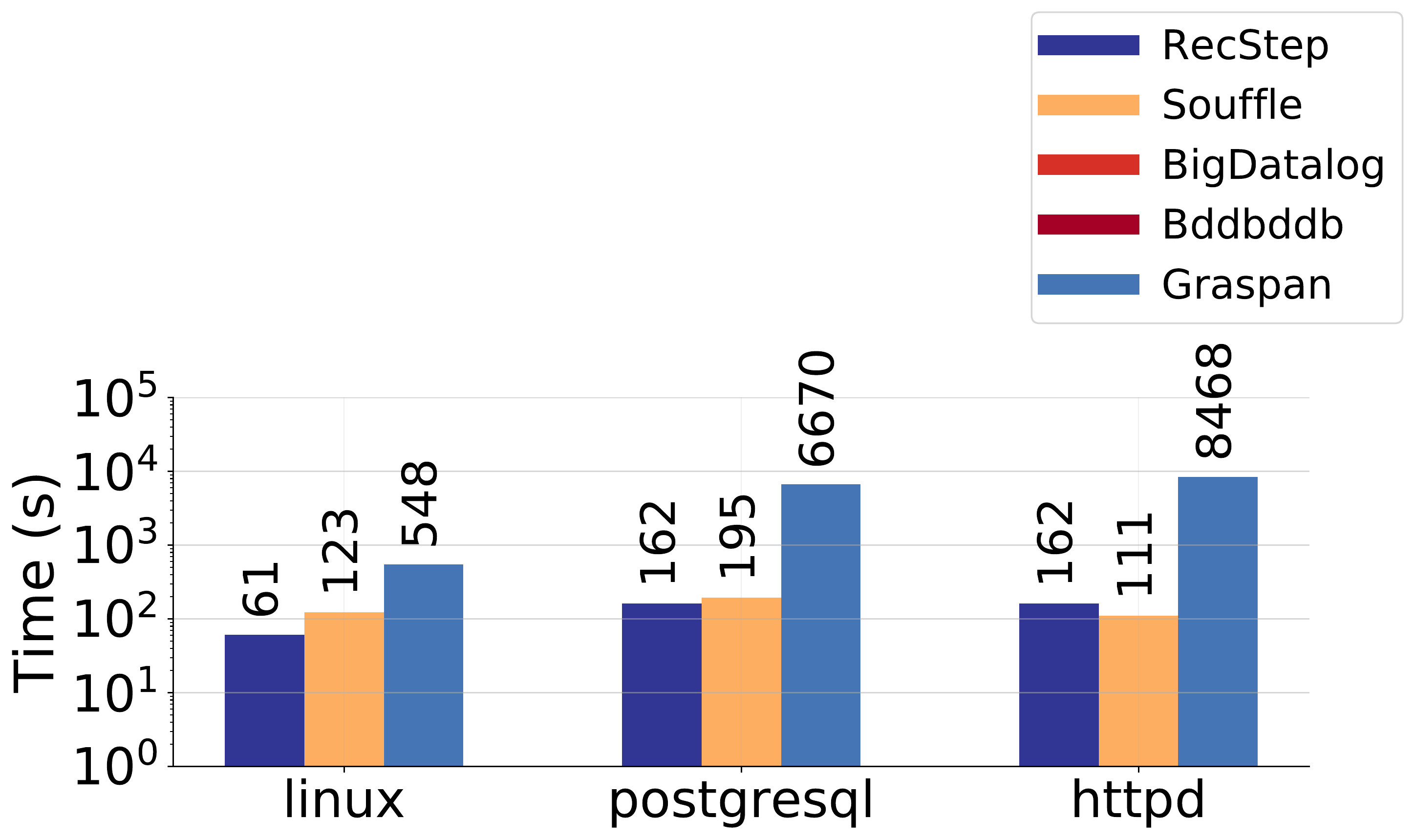}
        \caption{CSPA on systems programs}
        \label{fig:points_to}
    \end{subfigure}
    \caption{Performance Comparison on Program Analyses}
    \label{fig:program_analysis}
\end{figure*}

\introparagraph{Comparison With Other Systems.}
In this section, we report experimental results over our benchmarks for several other \datalog\ systems and Graspan.
For each \datalog\ program and dataset shown in the comparison results, we run the evaluation four times (with the exception of \bddbddb,
 since its runtime is substantially longer than all other systems across the workloads), we discard the first run and report the average time of the last three runs. For each system, we report the total execution time, including the time to load data from the disk and write data back to the disk. For BigDatalog, since its evaluation is \textit{parameter workload dependent} based on the available resources provided (e.g., memory), and its performance depends on both of the supplied parameters, datasets and the programs, we tried different combinations of parameters (e.g., different join types) and report the best runtime numbers.  For comparison purpose, we also display the results of BigDatalog that runs on the full cluster from \cite{BigDatalog} (Distributed-BigDatalog in Figure \ref{fig:tc_sg}, \ref{fig:rmat_exp}, \ref{fig:real_world_exp},  which has $15$ worker nodes with 120 CPU cores and 450GB memory in total).
	As we will see next, the experiments show that our system can efficiently evaluate \datalog\ programs for both large-scale graph analytics and program analyses, by being able to efficiently utilize the available resources on a single node equipped with powerful modern hardware (multi-core processors and large memory). Specific runtime numbers are not shown in Figure \ref{fig:tc_sg} and Figure \ref{fig:andersen_analysis} due to the space limit.

\introparagraph{TC and SG Experiments.}
For TC and SG, our system uses \textsc{pbme} as discussed in Section 5. 
Since the G$n$-$p$ graphs are very dense, in each iteration  intermediate results
of large sizes are produced. Hence, the original QuickStep operators run out of memory due to the high materialization cost and demand for memory. 
By using a \textit{bit-matrix} data structure, the evaluation naturally fuses the join and deduplication into a single computation, avoiding the materialization cost and heavy memory usage.
Our system is the only one that can complete the evaluation for TC and SG on all G$n$-$p$ graphs 
(the runtime bar is not shown if the system fails to finish the evaluation due to OOM or timeout ($>10h$) is observed ).
The evaluation time of all four systems
is shown in Figure~\ref{fig:tc_sg} and Figure~\ref{fig:tc_sg_memory} shows the memory consumption of each system.

\smallskip

For TC, except for Distributed-BigDatalog, our system outperforms all other systems on all G$n$-$k$ graphs (Distributed-BigDatalog is only slightly faster than \datalogsys\
on {\tt G20K}, {\tt G40K} and {\tt G80K}).  
For {\tt G5K}, {\tt G10K}, {\tt G10K-0.01}, and {\tt G10K-0.1}, our system even outperforms Distributed-BigDatalog, which uses 120 cores and 450GB memory. 
For graphs that have more number of vertices, Distributed-BigDatalog slightly outperforms our system due to the additional CPU cores and memory it uses for evaluation.

 Due to the use of \bdds,  \bddbddb can only efficiently evaluate graph analytics expressed in Datalog when the graph has a relatively small number of vertices and when the proper {variable ordering} is given.\footnote{The size of BDD is highly sensitive to the variable ordering used in the binary encoding;
finding the best ordering is NP-complete. We let \bddbddb pick the ordering.}
  When the evaluation violates either of these two conditions, \bddbddb is a few orders of magnitude slower than other systems as shown in graphs {\tt G5K}, {\tt G10K}, {\tt G10K-0.01}, {\tt G10K-0.1}. For graphs {\tt G20K}, {\tt G40K}, {\tt G80K}, \bddbddb runs out of time ( $> 10h$). Souffle runs out of memory when evaluating TC on {\tt G80K}.

\smallskip
 Compared to TC, the evaluation of SG is more memory demanding and computationally expensive as observed in Figure \ref{fig:sg} and \ref{fig:tc_sg_memory}. Except for our system, all other systems either use up the memory before the completion of the evaluation of SG or run into timeout ( $> 15h$) on some of the G$n$-$k$ graphs. Unlike TC, we observe that \datalogsys\ on SG evaluation is much more sensitive to the graph density. 

\begin{figure*}[t]
    \centering
    \begin{subfigure}[b]{0.3\textwidth}
        \centering
        \includegraphics[scale=0.16]{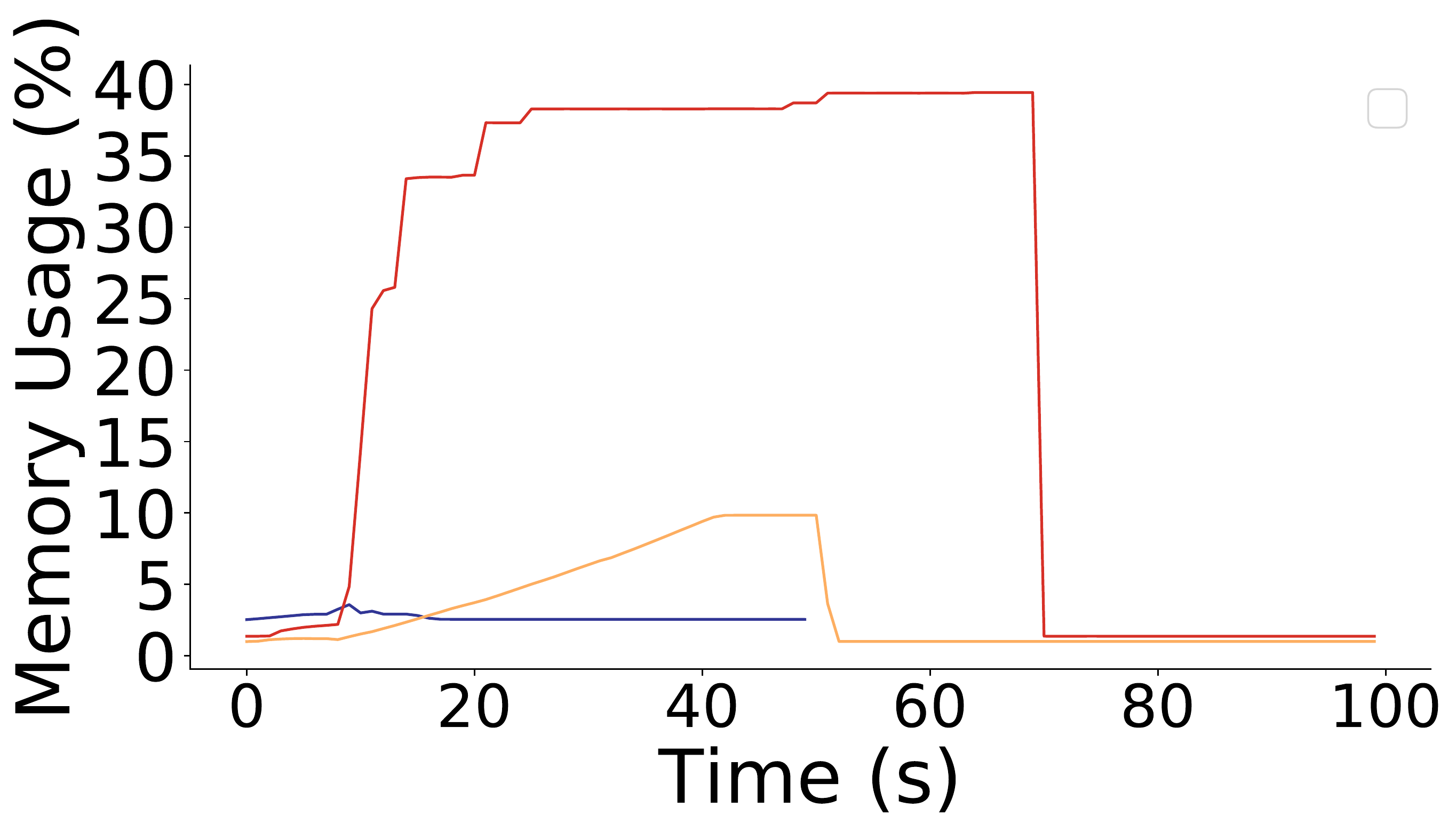}
        \caption{REACH}
        \label{fig:reach_memory}
    \end{subfigure}
    \begin{subfigure}[b]{0.3\textwidth}
        \centering
        \includegraphics[scale=0.16]{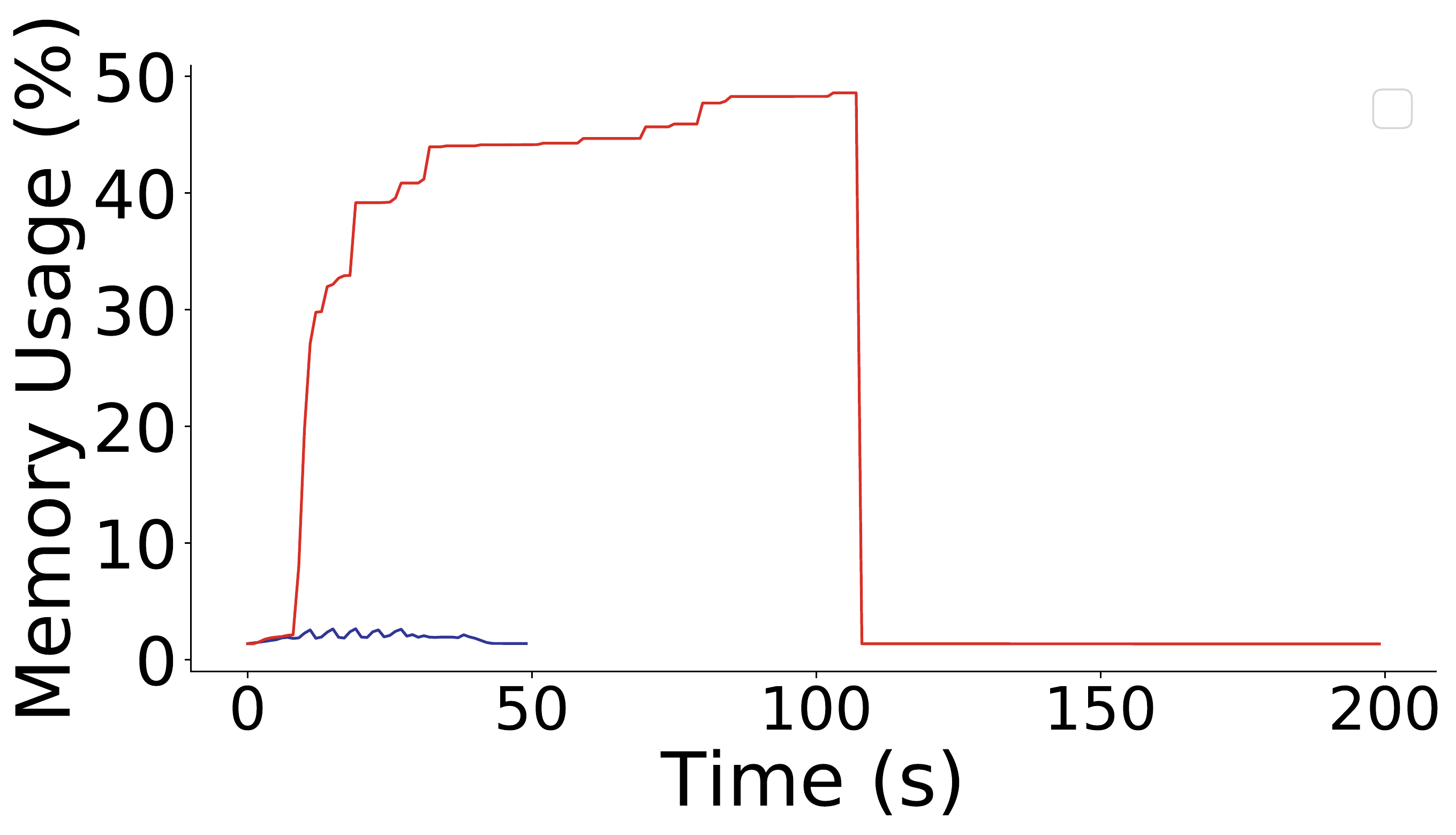}
        \caption{CC}
        \label{fig:cc_memory}
    \end{subfigure}
    \begin{subfigure}[b]{0.3\textwidth}
        \centering
        \includegraphics[scale=0.16]{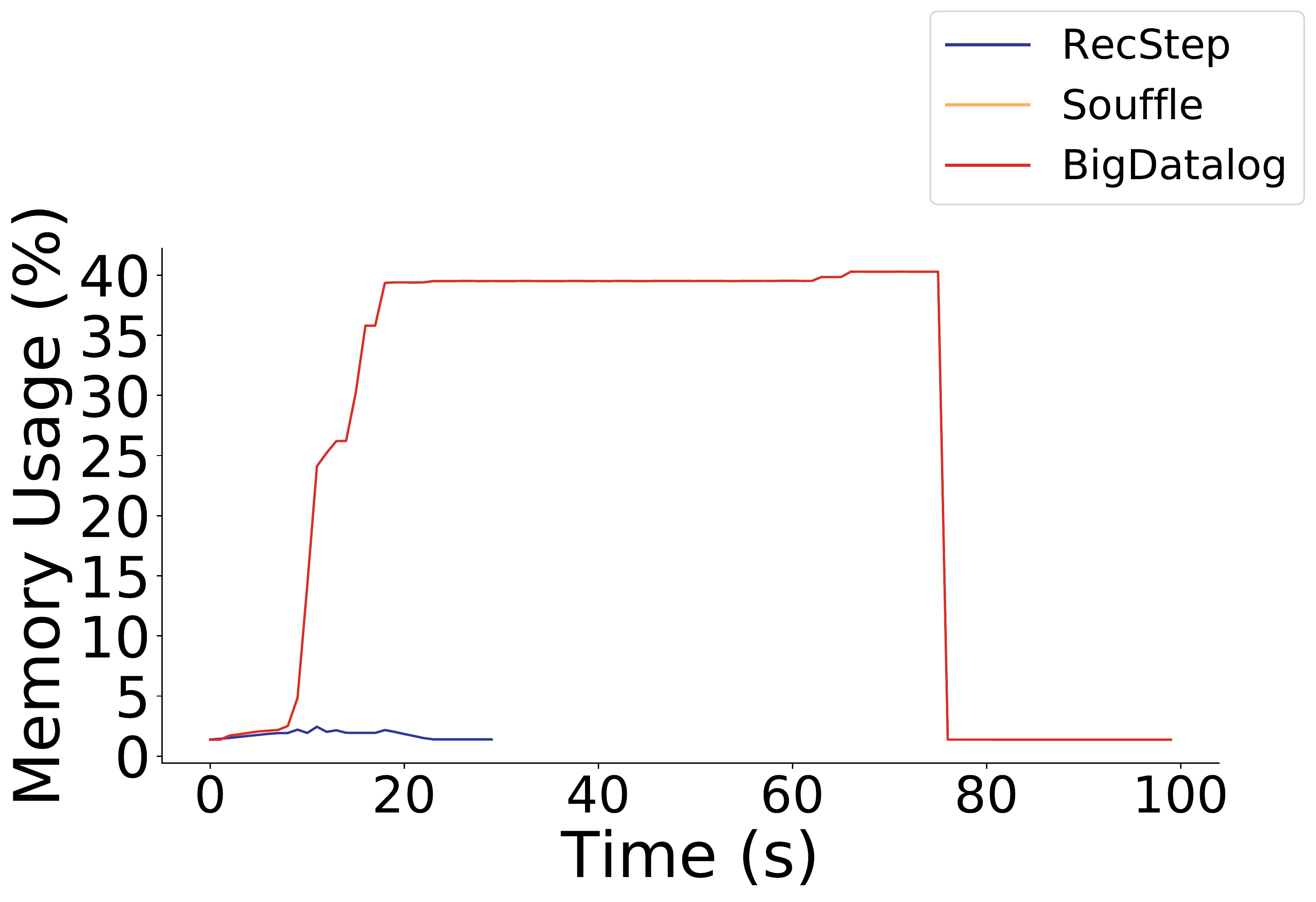}
        \caption{SSSP}
        \label{fig:sssp_memory}
    \end{subfigure}
    \caption{Memory Consumption on {\tt livejournal}.}
    \label{fig:livejournal_memory}
\end{figure*}

\begin{figure*}[!ht]
    \centering
    \begin{subfigure}[b]{0.3\textwidth}
        \centering
        \includegraphics[scale=0.16]{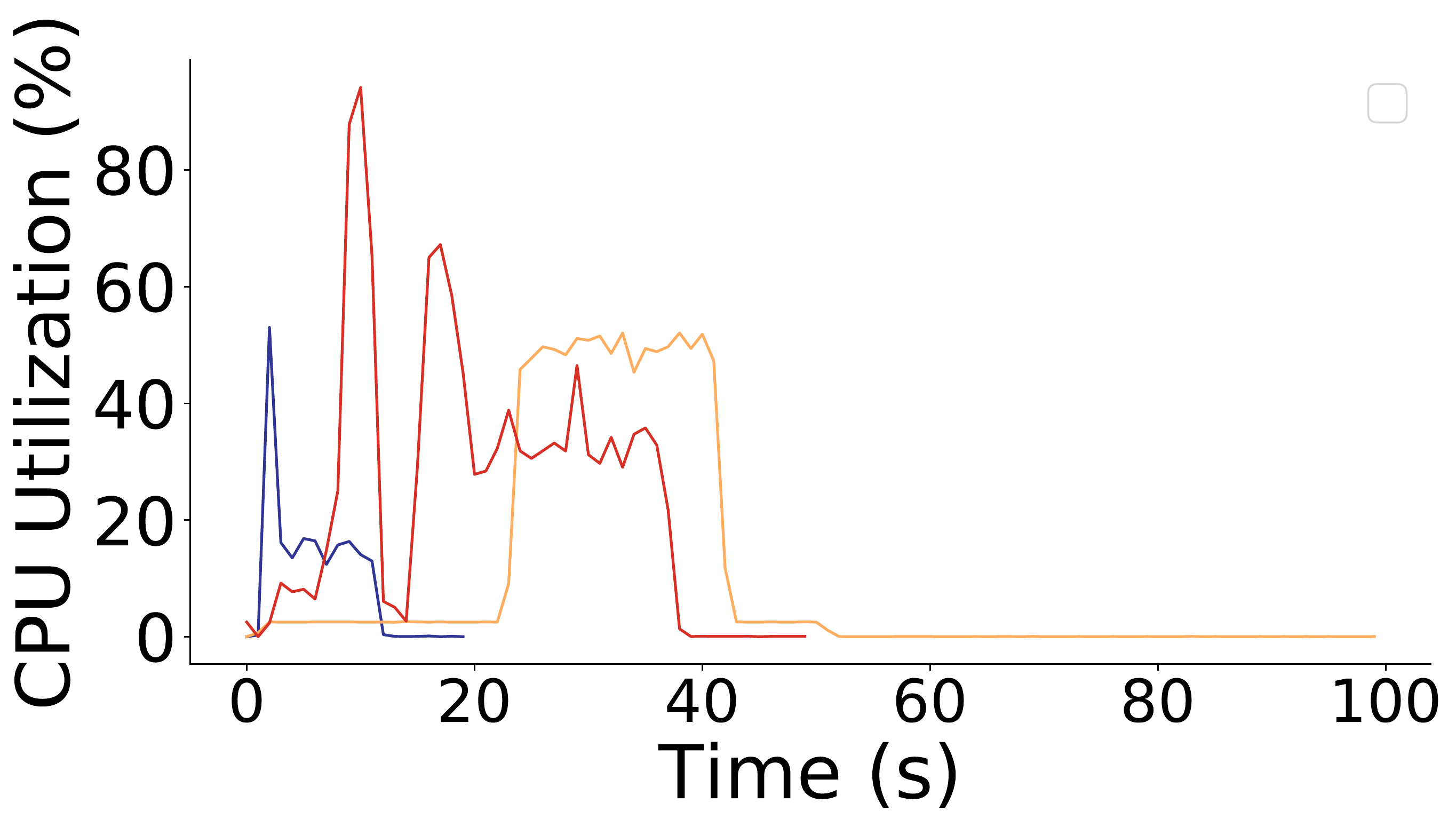}
        \caption{AA on {\tt dataset 5}}
        \label{fig:andersen_memory}
    \end{subfigure}
    \begin{subfigure}[b]{0.3\textwidth}
        \centering
        \includegraphics[scale=0.16]{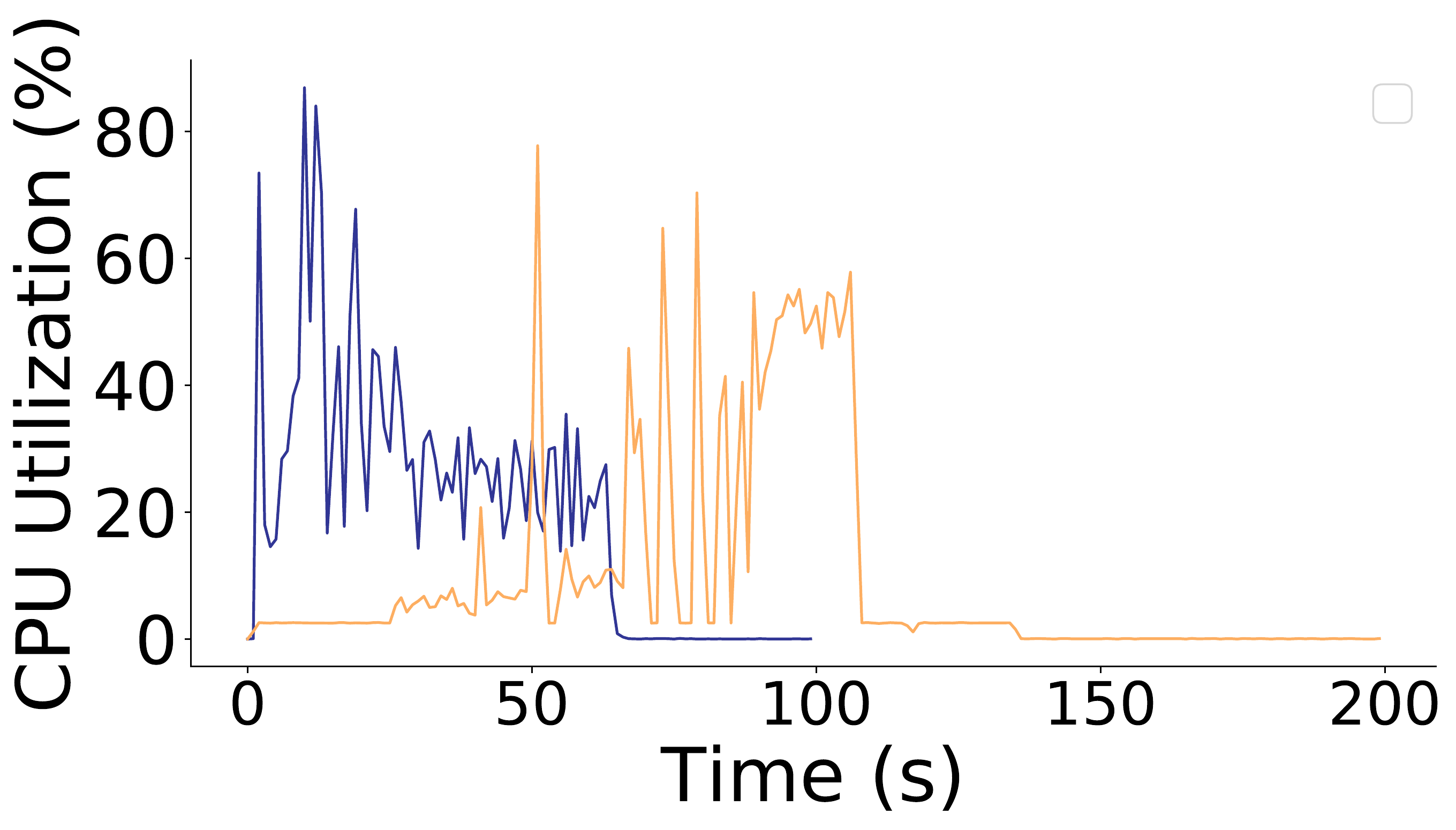}
        \caption{CSPA on {\tt linux}}
        \label{fig:dataflow_memory}
    \end{subfigure}
    \begin{subfigure}[b]{0.3\textwidth}
        \centering
        \includegraphics[scale=0.16]{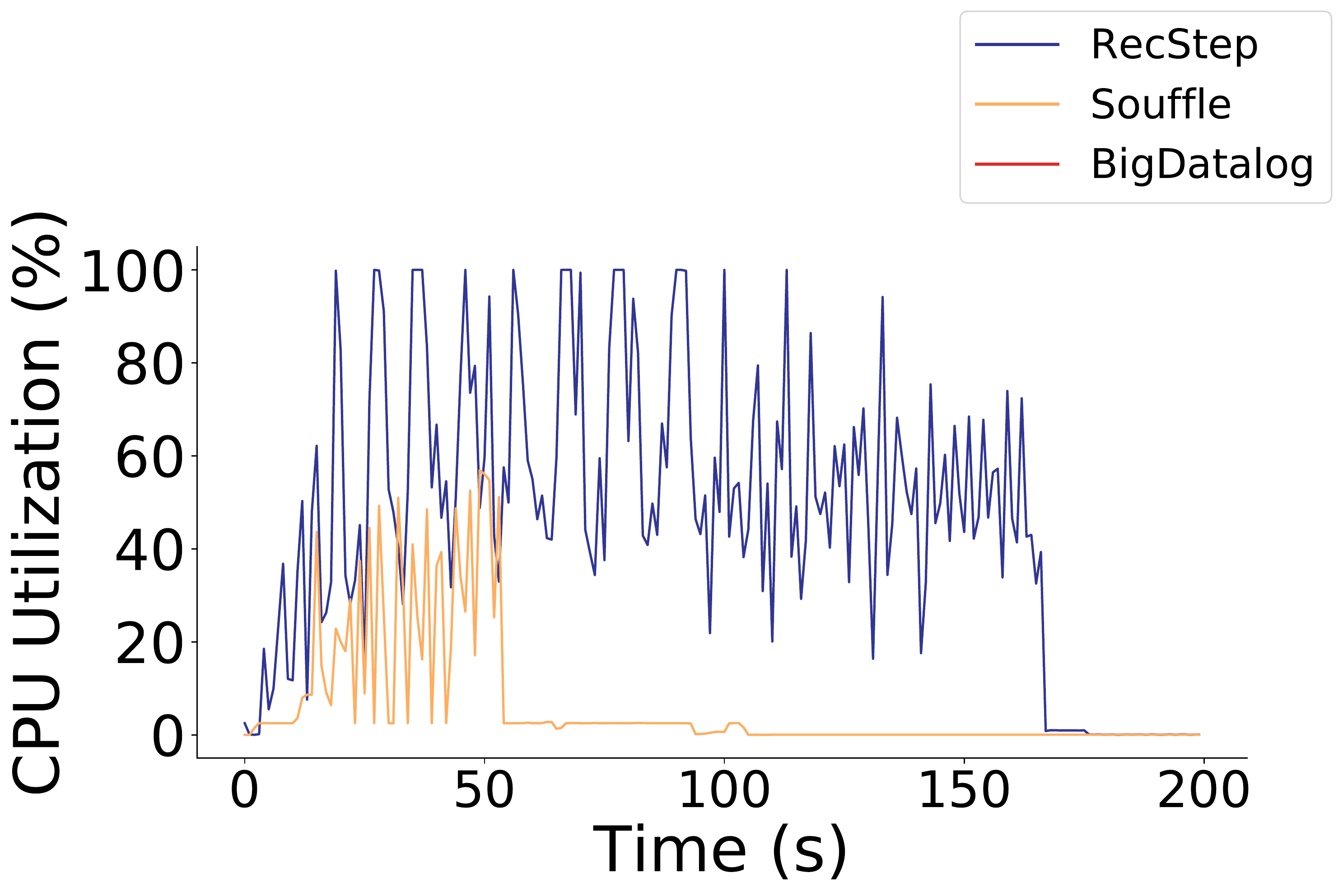}
        \caption{CSPA on {\tt httpd}}
        \label{fig:points_to_memory}
    \end{subfigure}
    \caption{ CPU Utilization on Program Analyses}
    \label{fig:program_analysis_cpu}
\end{figure*}

\introparagraph{Experiments of Other Graph Analytics}
Besides TC and SG, we also perform experiments running REACH, CC and SSSP on both the {\tt RMAT} graphs and the real world graphs (Table \ref{tab:benchmark}), comparing the execution time and memory consumption (on {\tt livejournal}) of our system with Souffle and BigDatalog (Figure~\ref{fig:real_world_exp}, \ref{fig:livejournal_memory}) . Since Soufflle does not support recursive aggregation (which shows in CC and SSSP), we only show the execution time results of our system and BigDatalog for CC and SSSP.  \bddbddb is excluded, since the number of vertices of all the graphs is too large.

As mentioned in \cite{BigDatalog}, the size of the intermediate results produced during the evaluation of REACH, CC, SSSP is $O(m)$, $O(dm)$ and $O(nm)$, where $n$ is the number of vertices, $m$ is the number of edges and $d$ is the diameter of the graph. For convenience of comparison, we follow the way in which \cite{BigDatalog} presents the experiments results: for REACH and SSSP, we report the average run time over ten randomly selected vertices. We only consider
an evaluation {complete} if the system is able to finish the evaluation on \textit{all ten vertices} for REACH and SSSP, otherwise, the evaluation is seen as \textit{failed} (due to OOM). The corresponding point of failed evaluation is not reported in Figure \ref{fig:rmat_exp} (on {\tt RMAT} graphs) and is shown as \textit{Out of Memory} in Figure \ref{fig:real_world_exp} (on real world graphs).

Besides Distributed-BigDatalog, \datalogsys\ is the {only} system that completes the evaluation for REACH, CC, SSSP on all {\tt RMAT} graphs and real world graphs, and is 3-6X faster than other systems using scale-up approach on all the workloads that other systems manage to finish (as shown in Figures~\ref{fig:rmat_exp} and \ref{fig:real_world_exp}); compared to Distributed-BigDatalog, \datalogsys\ shows comparable performance using far less computational resource.
Both BigDatalog and Souffle fail to finish evaluating some of the workloads due to OOM.
As shown in Figure \ref{fig:rmat_exp}, the execution time of our system increases nearly proportionally to the increasing size of the dataset on all three graph analytics tasks. In contrast, Souffle's parallel behavior is workload dependent  though it efficiently evaluates dataflow and points-to analysis (Fig \ref{fig:dataflow_analysis}, \ref{fig:points_to}),  it does not fully utilize all the CPU cores when evaluating REACH (Fig\ref{fig:reach_real_world}, \ref{fig:reach_rmat}) and Andersen's analysis (Fig \ref{fig:andersen_analysis} ) .The CPU utilization of different system evaluating Andersen's analysis, CSPA is visualized in Figure \ref{fig:program_analysis_cpu}.

\introparagraph{Program Analysis.}
We perform experiments on Andersen's analysis using the synthetic datasets (generated based on a real-world dataset).
Besides, we also conduct experiments comparing the execution time of CSPA and CSDA on the real system programs in \cite{Graspan}. Nonlinear recursive rules are commonly observed in \datalog\ programs for program analysis, and the results help us understand the behavior of our system and other systems when evaluating Datalog programs involving/without involving complex recursive rules.

\smallskip
For Andersen analysis, the number of variables (the size of active domains of each \edb\ relation) increases from {\tt dataset 1} to {\tt dataset 7}. Our system outperforms all other systems on every dataset. The performance of \bddbddb is comparable to other systems when the number of variables being considered is small ({\tt dataset 1} and {\tt dataset 2}), but the runtime increases a lot when the number of variables grows, due to its large overhead of building the \bdd. BigDatalog outperforms Souffle on large datasets, since Souffle does not parallelize the computation as effectively.

\smallskip
All three systems significantly outperform Graspan on both CSPA and CSDA, as shown in Figure \ref{fig:dataflow_analysis} and Figure \ref{fig:points_to}. Since BigDatalog does not support mutual recursion, it is not present in Figure \ref{fig:points_to}). The inefficiency of Graspan is mainly due to its frequent use of sorting, coordination during the computation and relatively poor utilization of multi-cores for parallel computation.

CSDA is the only \datalog\ program on which \datalogsys\ is outperformed by Souffle and BigDatalog and the reasons are two-fold. First, the evaluation of CSDA on all
three datasets needs many iterations ($\sim 1000$), and thus the overhead of triggering each query including compiling the query plan accumulates. There is also an additional overhead from the \textsf{analyze} calls and the materialization cost of the intermediate results. Compared to this overhead, the cost of the actual computation is much smaller. The second reason is that the rules in CSDA are simple and linear. Since the input data and the intermediate results produced in each iteration is small in size, the RDBMS cannot fully utilize the available cores.

In contrast, CSPA has more rules and involves {nonlinear recursion}, producing large $\Delta$ and intermediate results at each iteration. This enables \datalogsys\ to exploit both data-level and multiquery-level parallelism. Figure \ref{fig:points_to} shows the evaluation time for CSPA: while Souffle slightly outperforms our system on the {\tt httpd} dataset, \datalogsys\ outperforms Souffle and Graspan on the other two datasets.

\section{Conclusion}

In this paper, we presented the design and implementation of \datalogsys, a general-purpose, parallel, in-memory Datalog solver,
along with the experimental comparison results of existing techniques.
Specifically, we demonstrated how to implement an efficient, parallel Datalog solver atop a relational database.
To achieve high efficiency, we presented a series of algorithms, data structures, and optimizations, at the level of \datalog compilation to \sql and at the level of the underlying \rdbms.
Our results demonstrate the scalability of our approach, its applicability to a range of application domains, and its competitiveness with highly optimized and specialized Datalog solvers.
The experimental evaluation helps revealing the advantages and disadvantages of the existing techniques and guides the design and implementation of \datalogsys.

\balance

\bibliographystyle{abbrv}
\bibliography{references}

\begin{appendix}

\section{Cost Model of DSD}

\begin{algorithm}[ht]
\caption{One-Phase Set-Difference (OPSD)}
\label{alg:one-phase-set-diff}
\begin{algorithmic} [1]
\State \textbf{Input:} $R, R_{\delta}$ 
\State \textbf{Output: $\Delta{R} \gets R_{\delta}-R$}
\State $HS \gets \mathsf{buildHashTable}(R)$ \label{alg:one-phase-set-diff:build_ht}
\State $\Delta R \gets \mathsf{antiJoinProbe}(R_{\delta}, HS)$ \label{alg:one-phase-set-diff:delta}
\end{algorithmic}
\end{algorithm}

\begin{algorithm}[ht]
\caption{Two-Phase Set-Difference (TPSD)}
\label{alg:two-phase-set-diff}
\begin{algorithmic}[1]
\State \textbf{Input:} $R, R_{\delta}$ 
\State \textbf{Output: $\Delta{R} \gets R_{\delta}-R$}
\State $B \gets$ smallest of  $R_{\delta}, R$  \label{alg:two-phase-set-diff:compute_r_start}
\State $P \gets$ largest of  $R_{\delta}, R$
\State $HB \gets \mathsf{buildHashTable}(B)$
\State $r \gets \mathsf{joinProbe}(P, HB) $ \Comment{$r \gets  B \Join P$} \label{alg:two-phase-set-diff:compute_r_end}
\State $Hr \gets \mathsf{buildHashTable}(r) $
\State $\Delta R \gets \mathsf{antiJoinProbe}(R, Hr)$ \Comment{$\Delta{R} \gets R_{\delta}-r$}
\end{algorithmic}
\end{algorithm}

The cost of OPSD and TPSD: 
\vspace{0.5ex}
\\(1)
\begin{align*}
Cost & (OPSD) =  \\
& C_b \cdot |R| + C_p \cdot |R_{\delta}| \\
Cost & (TPSD)  = \\
  & C_b \cdot  (min(|R|, |R_{\delta}|) + |r|) +  
   C_p \cdot (max(|R|, |R_{\delta}|)  + |R_{\delta}|)
\end{align*}
Furthermore, the difference between the cost of OPSD and cost of TPSD is:
\vspace{0.5ex}
\\(2)
\begin{align*}
& Cost(OPSD) - Cost(TPSD) = \\ 
& C_b \cdot |R|  -
C_b \cdot min(|R|, |R_{\delta}|) - 
C_b \cdot |r| - C_p \cdot max(|R|, |R_{\delta}|)
\end{align*}
We can see from the cost model that the selection between the two set-difference algorithms depends on the sizes of $S$  and $R$ (in every iteration) and thus we discuss the two cases separately as follows: \\
\textit{If $|R| \leq |R_{\delta}|$:}
\vspace{0.5ex}
\\(3)
\begin{align*}
Cost(OPSD) - Cost(TPSD) =  - (C_b \cdot |r| + C_p \cdot |R_{\delta}| )   < 0
\end{align*}
\textit{If $|R| > |R_{\delta}|$:}
\vspace{0.5ex}
\\(4)
\begin{align*}
Cost(OPSD) -  Cost & (TPSD)  =  \\
& (C_b - C_p) \cdot |R|   - 
 C_b \cdot (|R_{\delta}| + |r|) 
\end{align*}
We can see that the selection between OPSD and TPSD for set-difference is not clear based on the above formulation when $|R| > |R_{\delta}|$, thus we use additional
parameters to help with the analysis. Denoting $C_b = \alpha C_p$, $|R| = \beta |R_{\delta}|$ and $ |R_{\delta}| = \mu |r|$, where $\alpha, \beta, \mu \in  \mathbb{R^{+}}$. Then we can rewrite the cost difference between OPSD and TPSD (when $|R| > |R_{\delta}|$) as:
\vspace{0.5ex}
\\(5)
\begin{align*}
Cost(OPSD) - Cost & (TPSD)
 =  \\
& \mu \cdot |r| \cdot C_p \cdot
[\beta(\alpha - 1) - (\alpha + \frac{\alpha}{\mu})]
\end{align*}
While we can know the accurate value of $\beta$ and $\alpha$, there is no way to know the accurate value of $\mu$ beforehand (before selecting between OPSD and TPSD).  However, we do know
that $|r| \leq |R_{\delta}|$ and thus $\mu \geq 1$ (when $|r| = 0$, the cost difference can be easily calculated from (4))  . Using this information, we can further derive the \textit{lower bound} of  the cost difference:
\vspace{0.5ex}
\\(6)
\begin{align*}
 \mu \cdot |r| \cdot C_p \cdot
[\beta(\alpha - 1) - (\alpha + \frac{\alpha}{\mu})]  \\
\geq \mu \cdot |r| \cdot C_p \cdot
[\beta(\alpha - 1) - 2\alpha]
\end{align*}
Thus if $\beta \geq \frac{2 \alpha}{\alpha - 1}$, then $Cost(OPSD) - Cost(TPSD) > 0$ and TPSD is used for set-difference computation. In a word, there is a clear guide for the selection of two set-difference algorithm when $\beta \in (0, 1] \cup [\frac{2 \alpha}{\alpha - 1}, +\infty ]$. Though when $\beta$ lies in $(1, \frac{2 \alpha}{\alpha - 1})$,  the cost difference between two set-difference
algorithm is not certain, we observe that in many of the workloads,  the value difference of $\mu$ in two \textit{consecutive} iterations is small, and thus a heuristic approach for choosing between OPSD and TPSD when $\beta$ lies in $(1, \frac{2 \alpha}{\alpha - 1})$ is to use the value of $\mu$ in previous iteration to approximate the value of $\mu$ in the current iteration.  The value of $\alpha$ is pre-computed from \textit{offline training}: perform $k$ join runs on $n$ table pairs $\{ \langle S_i, R_i \rangle$ $|$ $i = 1, 2, ... n\}$ of different sizes (ensure $|R_i| \leq |S_i|$, suggesting the hash table is always built on $R_i$). In $j$ join run on $\langle S_i, R_i \rangle$, denote the build cost as ${{B_i}_j}$ and the probe cost as ${{P_i}_j}$,  then:
\vspace{0.5ex}
\\(7)
\begin{align*}
\alpha = \sum_i^n{\frac{\sum_j^k{\frac{{B_i}_j \cdot |R_i|}{{P_i}_j \cdot |S_i|}}}{k}}
\end{align*}
The values of $k, n$ and table pairs  $\{ \langle S_i, R_i \rangle | i = 1, 2, ... n\}$ are pre-specified.

\section{CPU Efficiency}
The CPU efficiency is \textit{workload dependent}. A workload ($w$) is defined as a combination of a specific dataset ($d$) and a specific program ($p$). The runtime of a certain system ($s$)
on a workload $w$ consisting of $d$ and $p$ is denoted as $t^s_{w(p,d)}$. Denoting the number of CPU cores given for $s$ (which supports multi-core computation) as $n$, the CPU efficiency ($ce$)
is defined as $ce = \frac{1}{t^s_{w(p,d)} \cdot n}$. Intuitively, if $s$ can \textit{efficiently} utilize CPU given multiple-cores, the value of $t^s_{w(p,d)} \cdot n$ should be relatively small and $ce$ should be 
relatively large. Table \ref{tab:cpu_efficiency} presents the CPU efficiency of different systems on multiple chosen representative workloads.

\begin{table*}[t]
\begin{center}
\begin{tabularx}{\textwidth} {C{3cm} C{2cm} C{2cm} C{4cm} C{2cm} C{2cm}}
\toprule
\textbf{} &\  \textbf{Graspan}  &\ \textbf{BigDatalog} &\ \textbf{Distributed-BigDatalog} &\ \textbf{Souffle}  &\ \textbf{\datalogsys} \\

\textbf{TC ({\tt G20K})} &\ -  &\ 2.75e-04   &\ 4.39e-04  &\ 2.92e-04 &\ \textbf{1.12e-03} \\

\textbf{SG ({\tt G10K}))} &\ - &\ 7.18e-05  &\ 3.47e-04  &\ 5.41e-04 &\ \textbf{2.45e-03} \\

\textbf{REACH ({\tt orkut})} &\ -   &\ 1.92e-04  &\ 4.17e-04  &\ 3.52e-04 &\  \textbf{1.32e-03} \\

\textbf{CC ({\tt orkut})} &\ -   &\  2.17e-04 &\ 2.53e-04 &\ - &\ \textbf{5.81e-04} \\

\textbf{SSSP ({\tt orkut})} &\ - &\ 1.81e-04 &\ 2.14e-04  &\ - &\ \textbf{1.00e-03}  \\

\textbf{AA ({\tt dataset 7})} &\ - &\ 2.20e-04 &\ -  &\ 5.65e-05 &\ \textbf{7.65e-04} \\

\textbf{CSDA ({\tt linux})} &\ 2.22e-06 &\ 1.29e-04 &\ - &\ \textbf{2.05e-04} &\ 5.81e-05 \\

\textbf{CSPA ({\tt linux})} &\ 4.56e-05 &\ - &\ - &\ 2.03e-04 &\ \textbf{4.10e-04}\\

\bottomrule
\end{tabularx}
\caption{CPU Efficiency of Different Systems} 
\label{tab:cpu_efficiency}
\end{center}
\end{table*}

\section{Time and Space Effect Analysis of Optimizations}
Besides the empirical study of the time and memory effect of each optimization technique shown in Figure \ref{fig:op} and Figure \ref{fig:recstep_op_memory}, we discuss here about the time and space effect of each optimization in more detail.

\begin{packed_enum}
\item {\em Unified \idb Evaluation} (\uie): different rules and different subqueries inside each recursive rule evaluating the same \idb relation are issued as a single query \textit{without} introducing \textit{explicit coordination overhead} for multiple concurrent running subqueries while enabling better parallelism and pipelining.  For in-memory evaluation, \uie enables to share the hash-table(s) built on the same table(s), thus merely increases the peak memory consumption (\edb tables and \idb tables from previous iterations already reside in memory). 

\item {\em Optimization On the Fly} (\oof):  lightweight analytical queries are executed on data that is \textit{already in memory} at specified \textit{breakpoint} to collect necessary statistical data, which adds small runtime overheads without influencing the peak memory usage. The technique can result in great performance boost by making the relatively-long running query run right after the breakpoint choose the correct query plan. However, if there are too many after-breakpoint queries that are \textit{too short} (shorter than the analytical queries, which is unlikely to happen frequently), then \oof will slow down the whole evaluation performance.

\item {\em Dynamic Set Difference} (\dsd): the technique guides the choice of the set-difference algorithm that takes less time and adds negligible computational overheads from cost-model computation, which is barely a few mathematical operations. By exploiting the cost-model for DSD, table of smaller size is chosen to build the hash table on and thus the memory consumption is minimized. 

\item{\em Evaluation as One Single Transaction} (\eost): for in-memory evaluation, when memory is big enough for all the data to fit into, delaying the dirty pages produced during any of the intermediate operations until the end of evaluation avoids non-necessary I/O cost and has no influence on the memory usage.

\item{\em Fast Deduplication} (\fastDedup): the technique speed-ups the \datalog program evaluation by reducing the cost of hashing (using tuple itself as hash value) and chance of collision compared to the original parallel global separate chaining hash table. It achieves memory saving by eliminating the use of extra hash key and pointer to the tuple when the number of attributes of the tuple is small. \fastDedup can possibly increase the peak memory usage during the deduplication phase if the number of attributes of the tuples in the table to be deduplicated is large (when the memory consumed by the compact key is larger than the memory needed by the hash key and pointer).
\end{packed_enum}

\end{appendix}

\end{document}